%% file: ms.tex
\documentclass[apj]{emulateapj}
\usepackage{apjfonts}
\usepackage{graphicx}

\shorttitle{Metallicity Distribution Function of the Galactic Halo}
\shortauthors{An et~al.}

\begin{document}

\title{The Stellar Metallicity Distribution Function of the Galactic Halo from SDSS Photometry}

\author{Deokkeun An\altaffilmark{1},
Timothy C.\ Beers\altaffilmark{2,3},
Jennifer A.\ Johnson\altaffilmark{4},
Marc H.\ Pinsonneault\altaffilmark{4},
Young Sun Lee\altaffilmark{3,5,11},\\
Jo Bovy\altaffilmark{6,12},
\v{Z}eljko Ivezi\'{c}\altaffilmark{7},
Daniela Carollo\altaffilmark{8,9},
Matthew Newby\altaffilmark{10}
}
\altaffiltext{1}{Department of Science Education, Ewha Womans University,
Seoul 120-750, Republic of Korea; deokkeun@ewha.ac.kr.}
\altaffiltext{2}{National Optical Astronomy Observatory, Tucson, AZ 85719.}
\altaffiltext{3}{Department of Physics \& Astronomy and JINA: Joint Institute
for Nuclear Astrophysics, Michigan State University, E. Lansing, MI 48824.}
\altaffiltext{4}{Department of Astronomy, Ohio State University,
140 West 18th Avenue, Columbus, OH 43210.}
\altaffiltext{5}{Department of Astronomy, New Mexico State University,
Las Cruces, NM 88003.}
\altaffiltext{6}{Institute for Advanced Study, Einstein Drive, Princeton, NJ 08540, USA.}
\altaffiltext{7}{Department of Astronomy, University of Washington,
Box 351580, Seattle, WA 98195.}
\altaffiltext{8}{Macquarie University Research Centre in Astronomy,
Astrophysics \& Astrophotonics, Department of Physics \& Astronomy,
Macquarie University, NSW 2109 Australia.}
\altaffiltext{9}{INAF-Osservatorio Astronomico di Torino, Strada Osservatorio 20,
Pino Torinese, 10020, Torino, Italy.}
\altaffiltext{10}{Department of Physics, Applied Physics and Astronomy,
Rensselaer Polytechnic Institute Troy, NY 12180.}
\altaffiltext{11}{Tombaugh fellow.}
\altaffiltext{12}{Hubble fellow.}

\begin{abstract}
We explore the stellar metallicity distribution function of the Galactic
halo based on SDSS $ugriz$ photometry. A set of stellar isochrones is
calibrated using observations of several star clusters and validated by
comparisons with medium-resolution spectroscopic values over a wide range
of metal abundance. We estimate distances and metallicities for individual
main-sequence stars in the multiply scanned SDSS Stripe~82, at heliocentric
distances in the range $5 - 8$~kpc and $|b| > 35\arcdeg$, and find that
the {\it in situ} photometric metallicity distribution has a shape that
matches that of the kinematically-selected local halo stars from
\citeauthor{ryan:91b}. We also examine independent kinematic information
from proper-motion measurements for high Galactic latitude stars in our sample.
We find that stars with retrograde rotation in the rest frame of the Galaxy
are generally more metal poor than those exhibiting prograde rotation, which
is consistent with earlier arguments by \citeauthor{carollo:07} that the halo
system comprises at least two spatially overlapping components with differing
metallicity, kinematics, and spatial distributions. The observed
photometric metallicity distribution and that of \citeauthor{ryan:91b} can be
described by a simple chemical evolution model by \citeauthor{hartwick:76}
(or by a single Gaussian distribution); however, the suggestive
metallicity-kinematic correlation contradicts the basic assumption in this
model that the Milky Way halo consists primarily of a single stellar population.
When the observed metallicity distribution is deconvolved using two Gaussian
components with peaks at [Fe/H] $\approx -1.7$ and $-2.3$, the metal-poor
component accounts for $\sim20\%$--$35\%$ of the entire halo population in
this distance range.
\end{abstract}


\section{Introduction}

Knowledge of the nature of the stellar Galactic halo, which collectively
preserves a detailed record of our Galaxy's formation in the early universe
\citep[e.g.,][]{eggen:62,searle:78}, has expanded a great deal
in the past few years. With the advent of large photometric
surveys such as the Sloan Digital Sky Survey \citep[SDSS;][]{york:00,edr,dr1,
dr2,dr3,dr4,gunn:06,dr5,dr6,dr7,dr8}, as well as the massive spectroscopic follow-up efforts
that have come from SDSS, in particular, the Sloan Extension for Galactic
Understanding and Exploration \citep[SEGUE;][]{yanny:09}, and the recently
completed SEGUE-2 extension (Rockosi et al., in preparation),
the opportunity to collect a vast amount of information on the
nature of the Galactic halo has arrived.
Detailed quantitative comparisons of these observations with the spatial, kinematical,
and chemical space inferred from theoretical predictions of models for the
formation of the stellar halo through the hierarchical merging of higher-mass
subhalos and accretion of lower-mass subhalos \citep[e.g.,][]{bullock:05,
johnston:08,font:11,mccarthy:12,tissera:12} are now within reach.

Much attention has been paid to understanding the nature of the stellar halo(s),
in terms of the chemical and kinematical properties of its constituent stars. For example,
based on the medium-resolution SDSS spectroscopy of ``local'' halo stars
($d_{\rm helio} < 4$~kpc), \citet{carollo:07,carollo:10} argued that our Milky
Way stellar halo is a superposition of two overlapping systems, the inner and
outer halos, that are distinct in metallicity, kinematics, and spatial distributions
\citep[see also][]{beers:12}. Furthermore, \citet{nissen:10} showed that high
[$\alpha$/Fe], metal-poor halo stars in the solar neighborhood are mainly on
prograde orbits, while those with low [$\alpha$/Fe] abundances are preferentially
found on retrograde orbits. High [$\alpha$/Fe] halo stars are likely associated with
a dissipative component of the Galaxy that experienced a rapid chemical evolution,
while the low [$\alpha$/Fe] stars could be accreted components from dwarf galaxies
that had lower star-formation rates. For example, one could point toward low-mass
(low metallicity) dwarf-like galaxies (surviving examples of which might include
the ultra-faint dwarfs discovered by SDSS, e.g., Willman et al. 2005,
Belokurov et al. 2006a,b, Zucker et al. 2006, and many others since; see also
the discussion of Carollo et al. 2012 and Qian \& Wasserburg 2012)
as likely progenitors of the outer-halo stellar population of the Milky Way.

The key observational ingredient in testing and studying the dual nature of
the Galactic halo is the metal abundance for individual stars. In particular,
such a diversity of stellar populations in the halo could have been imprinted
on the metallicity distribution function (MDF) of halo stars.
Previous results on the halo MDFs have been based primarily on
spectroscopic surveys of likely halo stars selected from kinematically-biased
\citep[e.g.,][]{ryan:91b,carney:96} or kinematically-unbiased [e.g., the HK survey of
\citet{beers:85,beers:92}; the Hamburg/ESO survey described by
\citet{christlieb:03} and \citet{christlieb:08}] searches for metal-poor stars.
These approaches have the advantage of reasonably efficiently identifying stars
of the lowest metallicity, which are of great interest in their own right, but
they suffer from selection biases that can be difficult to quantify. Biases are
of particular importance if one seeks to understand the global properties of the
stellar halo, a necessary step in order to tell a coherent story of the assembly
and evolution of large spiral galaxies such as the Milky Way. In this regard,
large photometric surveys such as SDSS can be used to recover relatively
unbiased information on the nature of the Galactic halo.

Broadband photometry can provide reasonably accurate estimates of stellar
metallicities, temperatures, and distances, at least for main-sequence stars. In
the pioneering work to recover stellar metallicities from SDSS $ugriz$
photometry, \citet{ivezic:08} devised a method of using polynomial regressions
based on spectroscopic calibrations of de-reddened $u-g$ and $g-r$ colors
\citep[see also][]{peng:12}, which
is similar to the traditional method of relating UV excess with positions on the
Johnson $UBV$ diagrams \citep[e.g.,][]{carney:79}. The clear advantage of using
a photometric metallicity technique is the efficiency of estimating
metallicities for main-sequence stars, which are the most plentiful
and representative sample of individual stellar populations.

As recognized by \citet{ivezic:08}, the polynomial-based photometric
metallicity technique they developed becomes quite insensitive at lower
metallicity. They compared photometrically estimated metallicities with those
obtained for a subset of the same sample of stars with available
spectroscopic determinations, and found that the photometric [Fe/H] estimates
effectively saturate at [Fe/H]$_{\rm spec}\la -2$, and cannot be extended to
lower metallicities due to limits placed by the level of photometric errors
obtained by SDSS.

Although the power of photometric metallicity determinations from broadband
colors is fundamentally limited at low metallicities, metallicity estimates
can indeed be pushed down to below [Fe/H]$\sim-2.5$,
through the use of well-defined color-metallicity relations. For
example, in \citet{an:09b}, we used theoretical isochrones, calibrated using sets
of observed cluster data, to estimate photometric metallicities for a large
number of stars in the halo, based on SDSS $gri$ photometry. Although individual
metallicity estimates exhibit significant errors, due to the weak metallicity
sensitivity of $gri$ colors, it was shown that application of the calibrated
isochrones can extend the low metallicity limit down to at least
[Fe/H]$\sim-2.5$, based on a comparison with SDSS spectroscopic abundance
measurements.

In this paper, we extend our previous efforts on the calibration of stellar
isochrones to provide improved photometric metallicity estimates, taking into
account the full set of information from the SDSS $ugriz$ measurements
(\S~\ref{sec:method}). A newly calibrated set of isochrones is developed,
superseding the set discussed by \citet{an:09a}. Based on these new isochrones,
we re-explore application of the photometric metallicity estimation technique
[which is clearly independent of that discussed previously by \citet{ivezic:08}]
to samples of stars with available SDSS photometry (\S~\ref{sec:sample}),
in order to construct an unbiased MDF of the
Galactic halo (\S~\ref{sec:results}). Based on our derived photometric MDFs and a
limited investigation of available kinematic information, we show that simple
chemical evolution models that are assumed to apply to a single stellar
population provide inadequate descriptions of the nature of the Galactic halo.

\section{Photometric Metallicity Technique}\label{sec:method}

The key ingredient for photometric estimation of stellar metallicities (as
well as distances) is the use of well-established stellar color-magnitude relations
over a wide range of stellar atmospheric parameters. For this purpose, we adopted
the formalism in \citet{an:09a}, where we used a set of YREC \citep{sills:00}
stellar isochrones in the SDSS filter system calibrated against observed $ugriz$
(open and globular) cluster photometry \citep{an:08} to estimate the stellar
metallicities and distances of individual stars.

As an initial exploration of this technique, we used the SDSS $gri$
data to constrain (median) photometric metallicities (hereafter [Fe/H]$_{\rm phot}$)
for bulk stellar populations in the halo \citep{an:09b}. However, it was necessary
to include $u$-band measurements to tightly constrain photospheric metal abundances, and
therefore to construct a precise MDF. This has a consequence of limiting the
application of the photometric technique to nearby main-sequence stars, because
of the relatively shallow survey limit in the SDSS $u$ passband ($u=22.0$ at the
95\% detection repeatability for point sources).

In this section, we describe the development of empirical corrections to
color-$T_{\rm eff}$ relations for stellar isochrones in the native SDSS filter
system \citep{an:09a}, including the $u$ passband, and test the accuracy of
color-$T_{\rm eff}$ relations and photometric metallicities using spectroscopic
estimates from the SEGUE Stellar Parameter Pipeline \citep[SSPP;][]{lee:08a,
lee:08b,allende:08,smolinski:11}

\subsection{Background: Calibration of Isochrones}\label{sec:isochrone}

In the following analysis we employ the same set of underlying stellar interior
models and the same $\alpha$-element enhancement scheme as in \citet{an:09a},
motivated by the observed behavior of these elements among field dwarfs and cluster
stars from spectroscopic studies \citep[e.g.,][]{venn:04,kirby:08}:
[$\alpha$/Fe]$ = +0.4$ at [Fe/H]$ = -3.0$,
[$\alpha$/Fe]$ = +0.3$ at [Fe/H]$ = -2.0, -1.5, -1.0$,
[$\alpha$/Fe]$ = +0.2$ at [Fe/H]$ = -0.5$, and
[$\alpha$/Fe]$ = +0.0$ at [Fe/H]$ = -0.3, -0.2, -0.1, +0.0, +0.1, +0.2, +0.4$.
A linear interpolation was made in this metallicity grid to obtain isochrones
at intermediate [Fe/H] values. We adopted an age of $13$~Gyr for
$-3.0 \leq {\rm [Fe/H]} \leq -1.2$, and $4$~Gyr at $-0.3 \leq {\rm [Fe/H]} \leq +0.4$,
with a linear interpolation between these two metallicity ranges. As described
below, our strategy minimizes the application of our calibrated stellar models
to main-sequence turn-off stars, for which colors are most sensitive to the
adopted age. For this reason, our main results are insensitive to the age assumption,
but readers are cautioned if they apply our models to stellar populations with
different ages. The above age-metallicity-[$\alpha$/Fe] relations are adopted
throughout this paper.

Colors and magnitudes predicted by theoretical models typically do not agree
with those observed from star clusters. Therefore, we defined color-$T_{\rm
eff}$-[Fe/H] corrections to better match the data. In \citet{an:09a}, we used
M67 cluster data to define color-$T_{\rm eff}$ relations in $u\, -\, g$, $g\,
-\, r$, $g\, -\, i$, and $g\, -\, z$, at the cluster's metal abundance (${\rm
[Fe/H]} = 0.0$). In \citet{an:09b}, we adopted these correction factors in
$gri$ color indices at solar abundance, and used a linear ramp between
[Fe/H]$=-0.8$ and [Fe/H]$=0.0$. In this weighting scheme, empirical corrections
become zero at [Fe/H]$=-0.8$ and below, so metallicity estimation of metal-poor
stars are essentially those obtained from pure theoretical calculations, and are
not affected by the empirical color corrections. This choice was motivated by
the fact that the models are in satisfactorily agreement with data in $gri$
color indices for globular clusters \citep{an:09a}. Above solar abundance, we
applied the M67-based color corrections to all models, assuming that offsets in
colors between the model and data are independent of the metal abundance of a
star.

\subsection{Updates: Calibration of Isochrones}\label{sec:calibration}

In principle, cluster sequences alone can be used to directly estimate stellar distances
and metallicities, as long as there exist well-measured cluster-star samples over a wide
range of [Fe/H]. However, cluster data are often noisy (e.g., some cluster
fiducials are found to be essentially superimposed, even if their [Fe/H] values
are different), so we choose instead to be guided by theory, in order to infer
stellar colors at any given metallicity (and/or age).

In particular, updates on the empirical calibrations of models were necessary to
adjust model $u\, -\, g$ colors, which have a profound impact on the accuracy of
photometric metallicity estimates, to better match observed cluster fiducial
sequences over a wide range of metal abundances. Below we describe our adopted
methodology used to generate a new set of calibrated isochrones for all of the
$ugriz$ color indices.

We expanded upon the strategy of \citet{an:09a} for the color-$T_{\rm eff}$
calibration by the addition of cluster fiducial sequences in \citet{an:08} over a wide
range of metal abundances. Although our pure theoretical models predict colors
that are consistent with observed fiducial sequences within the total systematic
and random errors \citep{an:09a}, there still remains a small, but suggestive
systematic residual pattern of the color offsets over $T_{\rm eff}$ and [Fe/H].
In the updated calibration set, we have attempted to minimize these effects to better
constrain estimates of the stellar parameters.

In the color-$T_{\rm eff}$-[Fe/H] calibration, which is described in detail
below, we used cluster fiducial sequences for several globular and open clusters in
\citet{an:08}. These clusters are listed in Table~\ref{tab:tab2}, along with our
adopted values for the cluster parameters, which are the same as those used
in our earlier model comparisons \citep{an:08}.
The [Fe/H] and E($B\, -\, V$) estimates for globular clusters are
from \citet{kraft:03}, who used \ion{Fe}{2} lines from high-resolution spectra
to compare colors derived from high-resolution spectroscopic determinations
of $T_{\rm eff}$ with the observed colors of the same stars. Distances to the
globular clusters are all {\it Hipparcos}-based subdwarf fitting distances.
We adopted the \citet{carretta:00} distance estimates whenever they are available,
since they employed the same metallicity scale for both subdwarfs and cluster stars
in the subdwarf-fitting technique; otherwise, we adopted distances in \citet{kraft:03}.
For M67, we took the average reddening and metallicity estimates from high-resolution
spectroscopy in the literature \citep[][and references therein]{an:07b}, and adopted
a cluster distance estimated from an empirically calibrated set of isochrones in the
Johnson-Cousins-2MASS system \citep{an:07b}. For NGC~6791, we adopted the average
[Fe/H] from high-resolution spectroscopic studies \citep[see references in][]{an:09a}.
The cluster's reddening and distance estimates are based on the application of our
calibrated isochrones in the Johnson-Cousins-2MASS system (An et~al., in preparation).

We employed {\tt UberCal} (Uber-calibration) magnitudes \citep{padmanabhan:08} for
calibrating the cluster systems, instead of the ``Photometric
Telescope (PT)''-calibrated magnitudes (hereafter {\tt Photo} magnitudes).\footnote{
See also {\tt http://www.sdss3.org/dr8/algorithms/fluxcal.php}.}
Our original cluster sequences in \citet{an:08} were on the {\tt Photo} system, where the
standard SDSS photometric pipeline \citep{lupton:02} was used to define stellar
colors and magnitudes in SDSS. However, \citet{padmanabhan:08} later devised
a method of improving a relative photometric calibration error by using repeat
measurements in the overlapping fields of the survey. Since DR7, SDSS takes the
{\tt UberCal} magnitudes as the default magnitudes.

\input{tab1.tex}

We transformed the fiducial sequences in the {\tt Photo} system onto the {\tt
UberCal} system, by applying zero-point differences between the two systems in
the cluster flanking fields, where we derived photometric zero points for the
cluster fiducials \citep{an:08}. The differences between the two calibrations
are not alarmingly large for the fiducial sequences used in the current study.
The mean differences are $+0.003\pm0.014$ mag, $+0.003\pm0.015$ mag,
$-0.001\pm0.010$ mag, $-0.005 \pm0.028$ mag, and $-0.003\pm0.013$ mag in
$ugriz$, respectively, in the sense of {\tt UberCal} minus {\tt Photo}
magnitudes, for all the cluster fields considered in \citet{an:08}, except from
the few imaging stripes from which we could not retrieve {\tt UberCal} magnitudes.
The zero-point differences between {\tt UberCal} and {\tt Photo} systems
for several cluster fields used in this work are listed in Table~\ref{tab:tab1}.

\begin{figure}
\epsscale{1.05}
\plotone{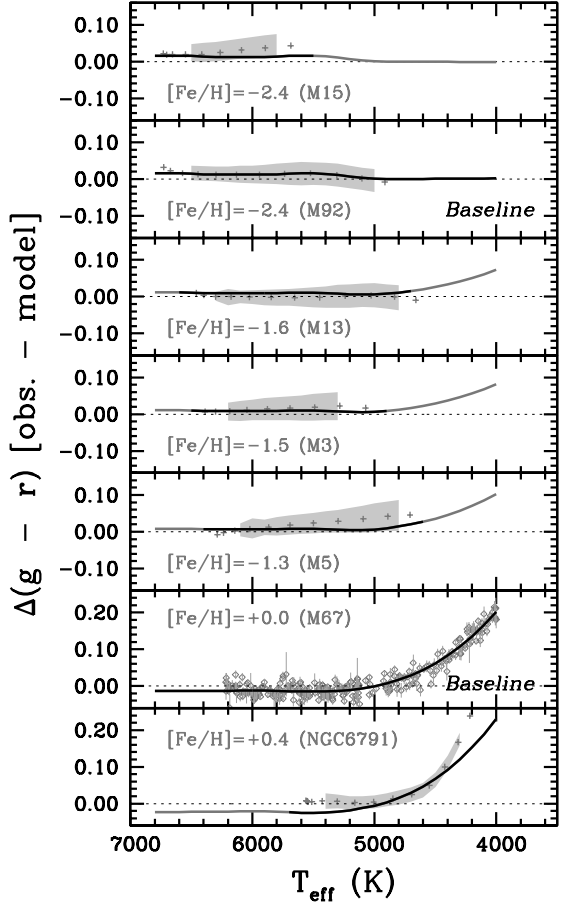}
\caption{Updated color-$T_{\rm eff}$-[Fe/H] calibrations of YREC isochrones.
Gray cross points are color differences in $g - r$ between the YREC models
and cluster fiducial sequences. A gray strip represents a $\pm1\sigma$ range of
a total systematic error in the comparison. Individual cluster cases are displayed
in increasing order of metallicity from the upper to lower panels. Solid lines
represent our derived color-$T_{\rm eff}$-[Fe/H] corrections, based on all of the
cluster comparisons shown above, where M92 and M67 comparisons are used as a baseline
(see text). Color differences for individual stars are shown for M67.
\label{fig:empcorr.gr.uber}}
\end{figure}

\begin{figure}
\epsscale{1.05}
\plotone{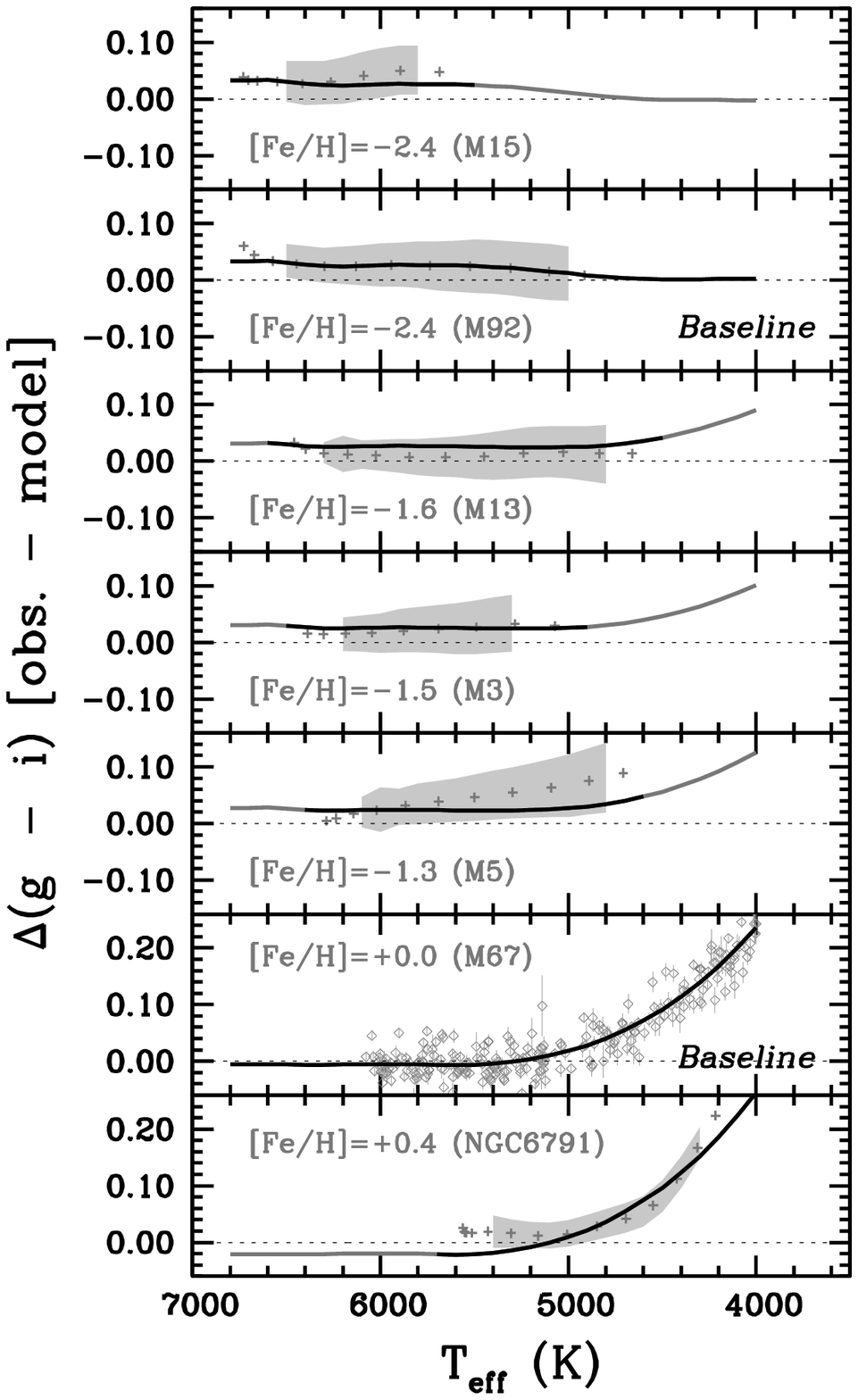}
\caption{Same as in Figure~\ref{fig:empcorr.gr.uber}, but for $g - i$.
\label{fig:empcorr.gi.uber}}
\end{figure}

\begin{figure}
\epsscale{1.05}
\plotone{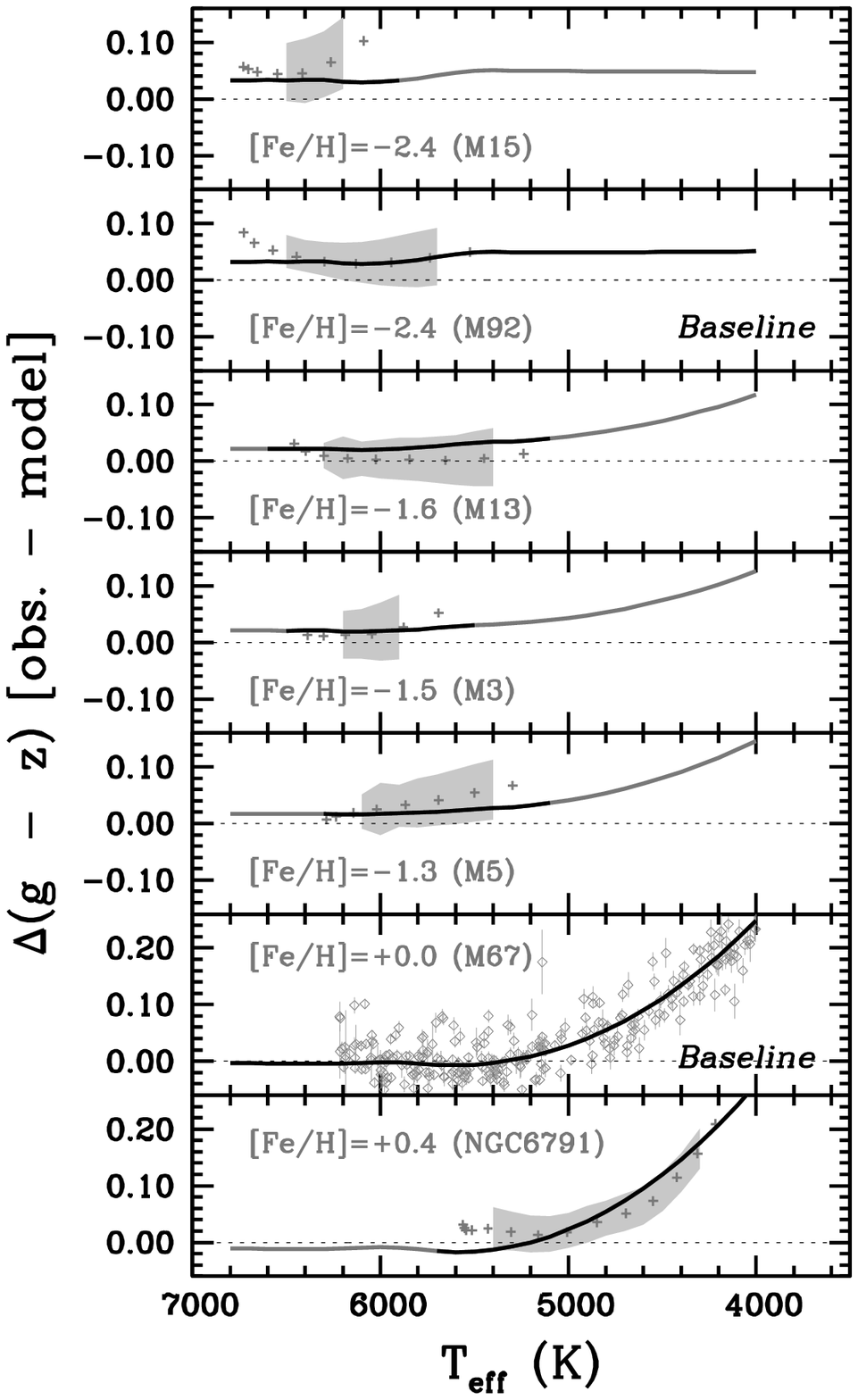}
\caption{Same as in Figure~\ref{fig:empcorr.gr.uber}, but for $g - z$.
\label{fig:empcorr.gz.uber}}
\end{figure}

\begin{figure}
\epsscale{1.05}
\plotone{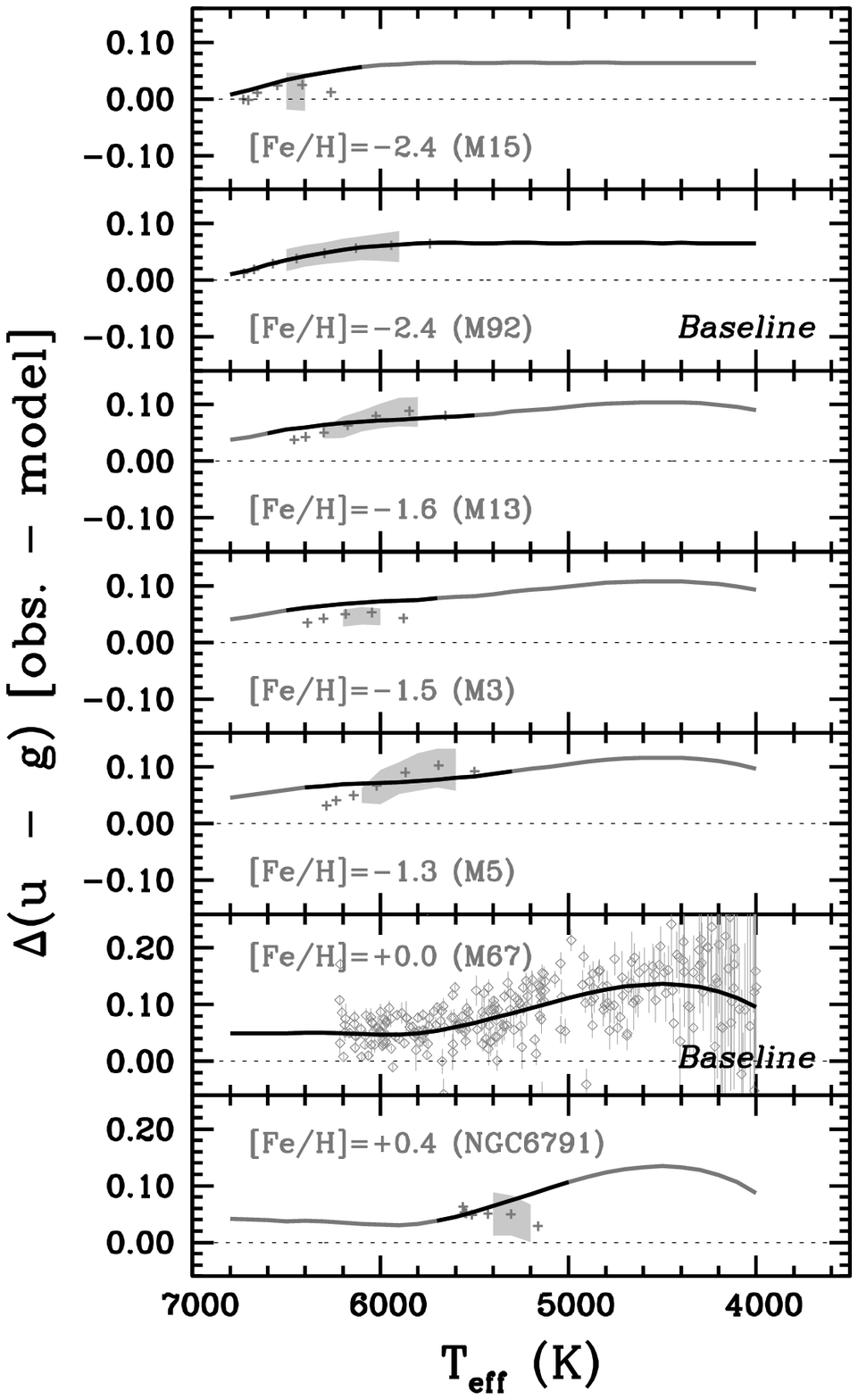}
\caption{Same as in Figure~\ref{fig:empcorr.gr.uber}, but for $u - g$.
\label{fig:empcorr.ug.uber}}
\end{figure}

Gray cross points in Figures~\ref{fig:empcorr.gr.uber}-\ref{fig:empcorr.ug.uber}
represent the color differences between cluster fiducial sequences and pure
theoretical stellar models in $g - r$, $g - i$, $g - z$, and $u - g$,
respectively. In each of the color indices, we have arranged the comparisons such
that comparisons for metal-poor clusters are shown in the upper panels, and those
for metal-rich clusters are shown in the lower panels. A gray strip represents
a $\pm1\sigma$ range of a total systematic error in the comparison, including
errors from the distance, reddening, age, and an assumed photometric zero-point
error on the fiducial sequences \citep[see also][for more details]{an:08}.

\input{tab2.tex}

Solid lines in Figures~\ref{fig:empcorr.gr.uber}--\ref{fig:empcorr.ug.uber} show
empirical color corrections derived using cluster fiducial sequences. We based
our calibration using cluster fiducial sequences for both M92 and M67. In other
words, color-$T_{\rm eff}$ relations were defined with the observed main sequences
of M92 at [Fe/H]$=-2.4$ \citep{kraft:03}, and those at [Fe/H]$=0.0$ were defined
with respect to M67 \citep[see][]{an:07b}. The choice of M92 and M67 for the base
case was due to the fact that these clusters are well-studied and have reliably
determined distances, foreground reddening, and metallicity estimates
\citep[][see also Table~\ref{tab:tab2}]{an:09a}. They both also have wide
$T_{\rm eff}$ coverage (see Figures~\ref{fig:empcorr.gr.uber}--\ref{fig:empcorr.ug.uber}).

To be consistent with our earlier color calibration with M67, we used the same color
corrections as in \citet{an:09a}. These are polynomial fits to the points over
$4000 \leq T_{\rm eff} {\rm (K)} \leq 6000$, and are expressed as follows:
\begin{eqnarray}
\Delta (g - r) &=& 7.610 - 19.910 \theta + 17.279 \theta^2 - 4.983 \theta^3,\label{eq:color1}\\
\Delta (g - i) &=& 6.541 - 16.313 \theta + 13.461 \theta^2 - 3.675 \theta^3,\\
\Delta (g - z) &=& 3.755 -  7.595 \theta +  4.535 \theta^2 - 0.672 \theta^3,\\
\Delta (u - g) &=&-7.923 + 24.131 \theta - 23.727 \theta^2 + 7.627 \theta^3,\label{eq:color2}
\end{eqnarray}
where $\theta \equiv T_{\rm eff}/5040$~K. These corrections are shown as solid
lines in Figures~\ref{fig:empcorr.gr.uber}--\ref{fig:empcorr.ug.uber} (second panels
from the bottom), and are in the sense that the above values should be added to
the model colors. For M92, we used the color differences between the model and
the cluster fiducial sequences as the baseline of the model calibration at the
cluster's metal abundance (see second panels from the top in
Figures~\ref{fig:empcorr.gr.uber}--\ref{fig:empcorr.ug.uber}).

A quadratic relation in [Fe/H] was then used to parameterize the metallicity
dependence of the color-temperature corrections, assuming the following functional form:
\begin{equation}
   \Delta_i (T_{\rm eff},{\rm [Fe/H]})
   = \delta_i (T_{\rm eff}) + \zeta_i {\rm [Fe/H]} + \xi_i {\rm [Fe/H]}^2,
\label{eq:calib}
\end{equation}
where $\Delta_i$ represents color corrections in each of $g - r$, $g - i$, $g - z$,
or $u - g$, while $\delta_i$ represents color-$T_{\rm eff}$ corrections in
Equations~(\ref{eq:color1})--(\ref{eq:color2}) in the $i^{th}$ color index at
solar metallicity. Here, $\zeta_i$ and $\xi_i$ are coefficients to be derived
from the fit, where we used fiducial sequences for M15, M13, M3, M5, and NGC~6791.
In fact, the problem is reduced
to a single parameter fit, since our models were defined to match the observed
main sequences of M67 and M92 at their metal abundances. The solid curves in
Figures~\ref{fig:empcorr.gr.uber}--\ref{fig:empcorr.ug.uber} show the resulting
color corrections on the color-$T_{\rm eff}$-[Fe/H] space.
We extrapolated color corrections below [Fe/H]$=-2.4$ using the best-fitting
parameters obtained from the above equation.

The $T_{\rm eff}$ vs.\ [Fe/H] range covered by our calibrating sample
clusters is sparse and non-uniform. In particular, hot stars
($T_{\rm eff} \ga 6000$~K) are not covered at or near solar abundance, while
cool, metal-poor stars ($T_{\rm eff} \la 5000$~K) are outside of our calibration
range. This is mainly because the clusters are found at varying distances from
the Sun, while all of the cluster images were taken in drift-scan or
time-delay-and-integrate (TDI) mode, with the same effective exposure time
of $54.1$ seconds per band. As a result, turn-off stars in M67 were too bright
and lower main-sequence stars in M92 were too faint for SDSS to obtain reliable
magnitudes. The cluster age is an additional factor that affects the non-uniform
coverage in the stellar parameter space.

In order to obtain full coverage of the color corrections in the $T_{\rm eff}$
vs.\ [Fe/H] plane, we assumed that the color difference from the model for
M67/M92 at the high/low $T_{\rm eff}$ end extends and remains constant to
$T_{\rm eff} = 7000$~K/$4000$~K, respectively. We note that this reasonable, but
arbitrary, extrapolation of the color-$T_{\rm eff}$ relations has minimal impact on
the results of our analysis, because our calibrating cluster sample is in fact
representative of the majority of stars detected in SDSS (in terms of $T_{\rm
eff}$ and [Fe/H], among other parameters). The purpose of the extrapolation is
to obtain stable color-$T_{\rm eff}$-[Fe/H] relations going forward.
Note that our model calibration is valid for main-sequence stars only.

Our adopted age of $13$~Gyr at $-3.0 \leq {\rm [Fe/H]} \leq -1.2$ in the
color calibration (\S~\ref{sec:isochrone}) is justified by our earlier
result that the main-sequence turn-off ages of our calibrating clusters are
approximately $13$~Gyr, when pure theoretical YREC models are directly used
in the age estimation \citep[see Table~9 in][]{an:09a}. To a first approximation,
a systematic error in the adopted age in the calibration results in a scale
error in our color-$T_{\rm eff}$ corrections. Therefore, special attention
should be paid when applying the models to stellar populations with different
age-metallicity relations.

\subsection{Stellar Parameter Search}\label{sec:param}

We applied calibrated stellar isochrones to the observed $ugriz$ magnitudes, and searched
for the best-fitting stellar parameters --- $T_{\rm eff}$ (or stellar mass), [Fe/H], and
an absolute magnitude (or distance) --- by minimizing the $\chi^2$ of the fit, defined as
\begin{equation}
\label{eq:chi}
\chi^2 = \sum_i \frac{(X_{{\rm obs},i} - X_{{\rm model},i})^2}{\sigma_i^2},
\end{equation}
for each star over $-3 \leq {\rm [Fe/H]} \leq +0.4$. Here, $X_{{\rm obs},i}$ and
$X_{{\rm model},i}$ are the observed and model magnitudes, respectively, in the
$i^{th}$ bandpass. Stellar mass, [Fe/H], and distances were set as free parameters
(i.e., three parameters and five data points for each star). This is equivalent to
fitting a model to the observed spectral energy distribution in the wavelength
versus flux space.

We adopted foreground dust estimates by \citet{schlegel:98}, with theoretical
extinction coefficients given by \citet{an:09a}:
\begin{equation}
R_\lambda {\rm (YREC)} = \frac{A_\lambda}{E(B\, -\, V)} = [4.858,3.708,2.709,2.083,1.513]
\end{equation}
where $\lambda=u,g,r,i,z$, respectively.
These extinction coefficients were computed using theoretical spectral energy distributions
and the standard \citet{cardelli:89} extinction curve. Our values are in good agreement with
those provided in \citet{girardi:04}.
The above values lie in between the default extinction coefficients in SDSS
\citep[][SFD98]{schlegel:98} and those in \citet[][SF11]{schlafly:11}:
\begin{equation}
R_\lambda {\rm (SFD98)} = [5.155,3.793,2.751,2.086,1.479],
\end{equation}
\begin{equation}
R_\lambda {\rm (SF11)}= [4.239,3.303,2.285,1.698,1.263].
\end{equation}

In terms of the $u\, -\, g$ colors, which have a significant impact on photometric
metallicity estimates, our $E(u\, -\, g) \equiv A_u - A_g$ coefficient is $18\%$ smaller
than the SFD98 value, but $19\%$ larger than the \citet{schlafly:11} value.
These differences are clearly important, and we discuss their effect on our
derived photometric MDFs in \S~\ref{sec:results} below.

\subsection{Comparisons with the SSPP Spectroscopic Estimates}\label{sec:sspp}

Photometric temperatures and metallicities in our approach are derived
simultaneously, based on the observed $ugriz$ photometry. Each of these
parameters are primarily constrained by different portions of the stellar spectral
energy distribution; $u\, -\, g$ colors are mostly responsible for the photometric
metallicities, while $griz$ colors are sensitive to temperatures. Nevertheless,
there is a moderate level of correlation between them, so checking a photometric
temperature scale is an important step toward obtaining accurate photometric metal
abundances of stars.

In \citet{pinsono:12}, we verified the accuracy of our photometric $T_{\rm eff}$
estimation procedure using the most recent temperature scale from the Infrared
Flux Method (IRFM) by \citet{casagrande:10}. Photometric temperatures were
estimated from $griz$ photometry in the {\it Kepler} Input Catalog
\citep{brown:11} using a calibrated set of
isochrones -- the same set of models used in the current work -- then they were
compared to the IRFM temperatures derived from 2MASS $J\, -\, K_s$ colors.
Overall, we found good agreement between these two fundamental temperature
scales ($\langle \Delta T_{\rm eff} \rangle \la \pm40$~K) at $4400$~K $< T_{\rm eff} < 6000$~K,
where the IRFM scale is well-defined; see \citet{pinsono:12} for detailed
discussions on this comparison.

\begin{figure}
\centering
\includegraphics[scale=0.65]{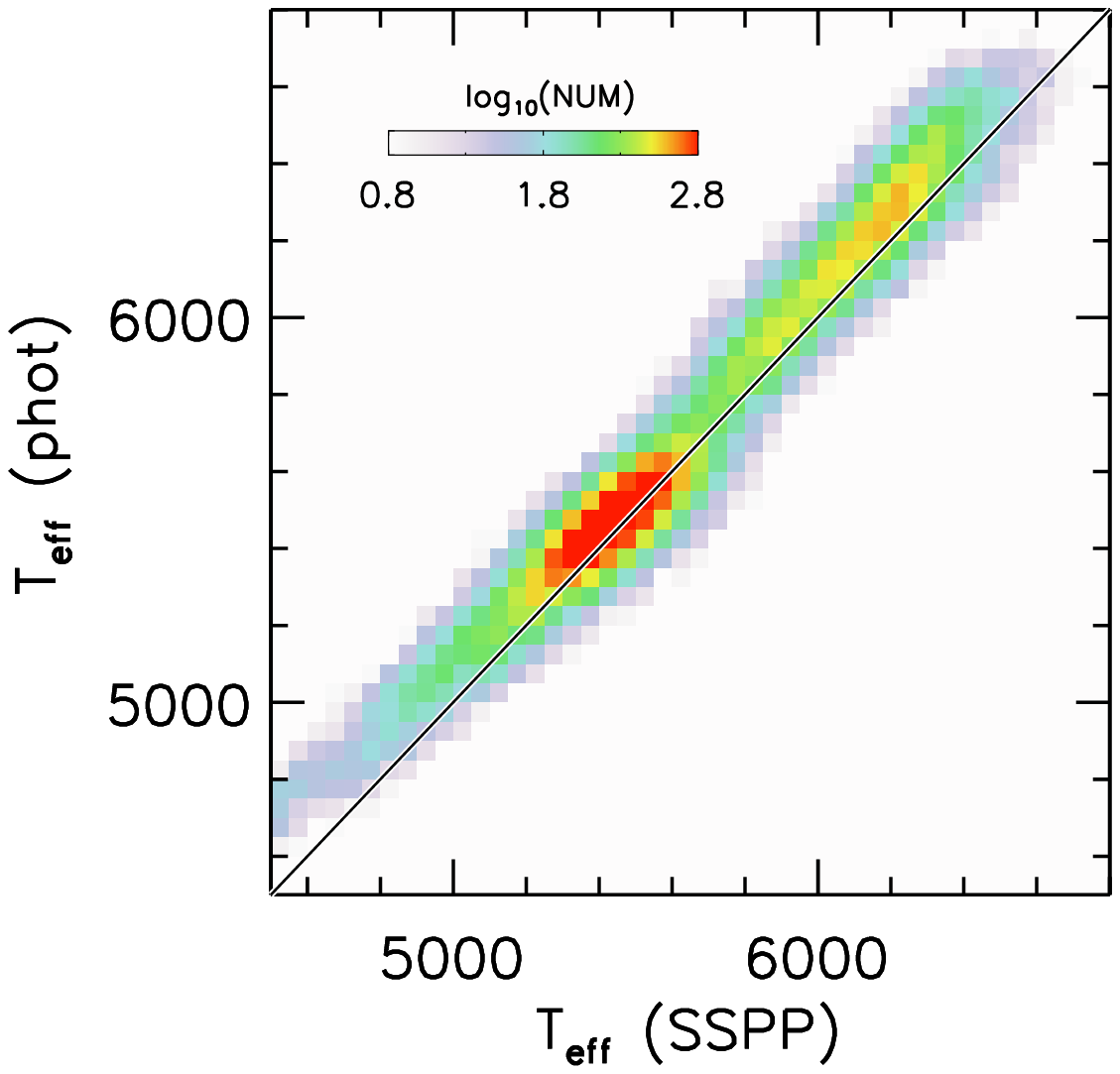}
\includegraphics[scale=0.65]{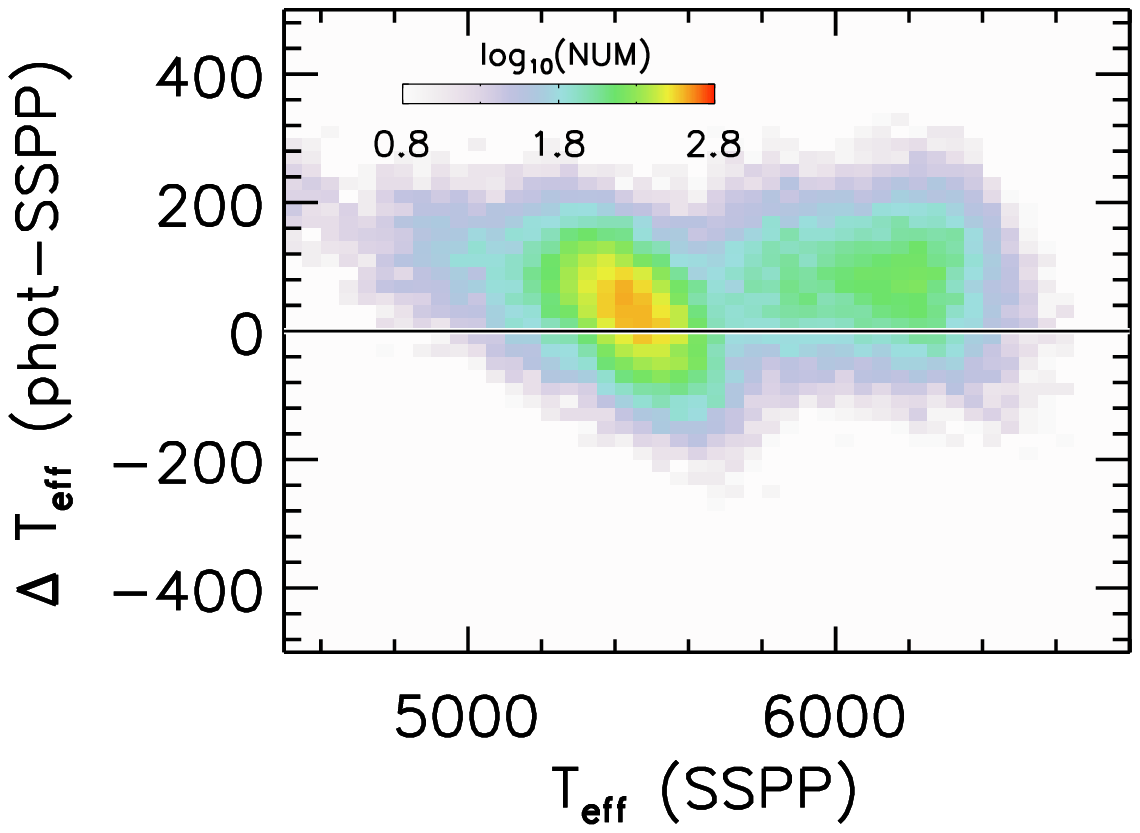}
\caption{Comparison between photometric and spectroscopic (SSPP) temperature estimates
for the SEGUE sample. The photometric estimates are based on all of the $ugriz$ photometry.
\label{fig:sspp.comp.teff}}
\end{figure}

Here we compare our photometric temperature estimates with the spectroscopic
estimates obtained from medium-resolution SDSS/SEGUE spectra \citep{yanny:09};
comparisons are shown in Figure~\ref{fig:sspp.comp.teff}. In this comparison we included
$51,999$ stars from the intial sample of $162,645$ objects that satisfy the following
selection criteria:
\begin{itemize}
  \item Sources are detected in all five bandpasses
  \item $\chi^2_{\rm min}/\nu < 3.0$, where $\chi^2_{\rm min}$ is a minimum
        $\chi^2$ defined in Equation~(\ref{eq:chi}) and $\nu$ ($=2$) is a degree of freedom
  \item $\sigma_u < 0.03$~mag
  \item $\log{g}_{\rm (YREC)} \geq 4.15$
  \item ${\rm S/N}_{\rm (SSPP)} > 20/1$
  \item $\log{g}_{\rm (SSPP)} \geq 3.5$,
\end{itemize}
where the sbuscript YREC and SSPP indicate parameters estimated from
the model isochrones and SSPP, respectively.

The first two criteria above require a good fit to the data, and the limit on the
$u$-band measurement errors exclude data points with less well-defined
photometric metallicity estimates. The $\log{g}$ limit restricts the analysis
to main-sequence dwarfs.
The last two criteria select dwarfs with reliable spectroscopic abundance
measurements. We applied further cuts, based on colors ($g\, -\, r > 0.2$) and
magnitudes ($r > 14$~mag), in order to remove stars with extremely blue
colors or saturated brightness measurements.
Most of the stars in the original SSPP sample were rejected due to their
large $u$-band errors. Even if a more conservative cut on
$\chi^2_{\rm min}/\nu < 10$ were used, there would still remain a total of
$58,335$ stars in the comparison.

As shown in Figure~\ref{fig:sspp.comp.teff}, there is a constant
$T_{\rm eff}$ offset between the photometric and SSPP temperature scales over
$5000$~K $< T_{\rm eff} < 6500$~K, where the maximum deviation is less than
$100$~K in this temperature range. However, the good agreement between our
temperature scale and the IRFM scale \citep{pinsono:12} ensures that our
temperature estimates are more or less closer to the fundamental temperature
scale.\footnote{Further note that $T_{\rm eff}$ is a defined quantity in $L = 4
\pi R^2 \sigma T_{\rm eff}$.} The random scatter seen in Figure~\ref{fig:sspp.comp.teff}
is $\sim100$~K for individual $T_{\rm eff}$ estimates.

\begin{figure}
\centering
\includegraphics[scale=0.65]{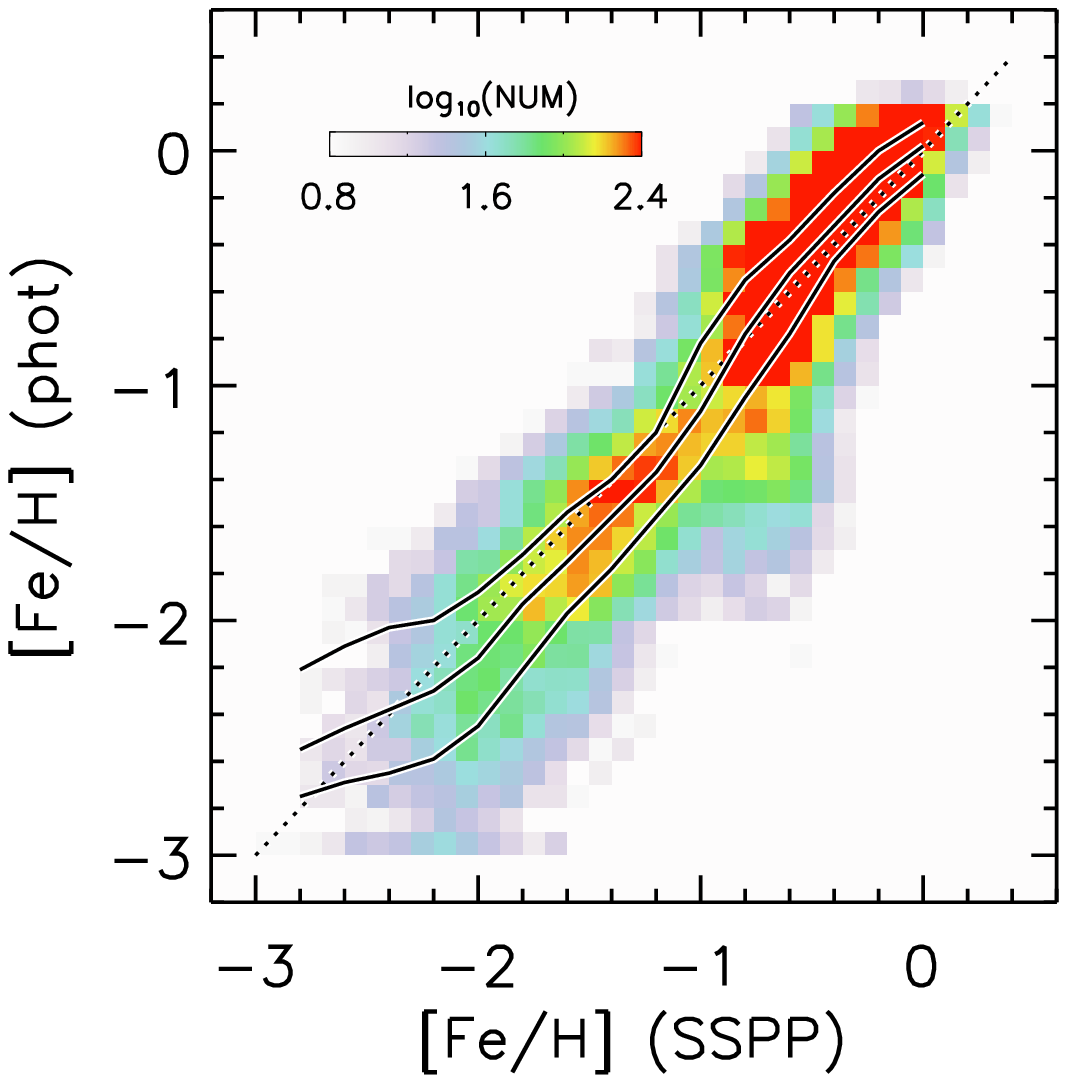}
\includegraphics[scale=0.65]{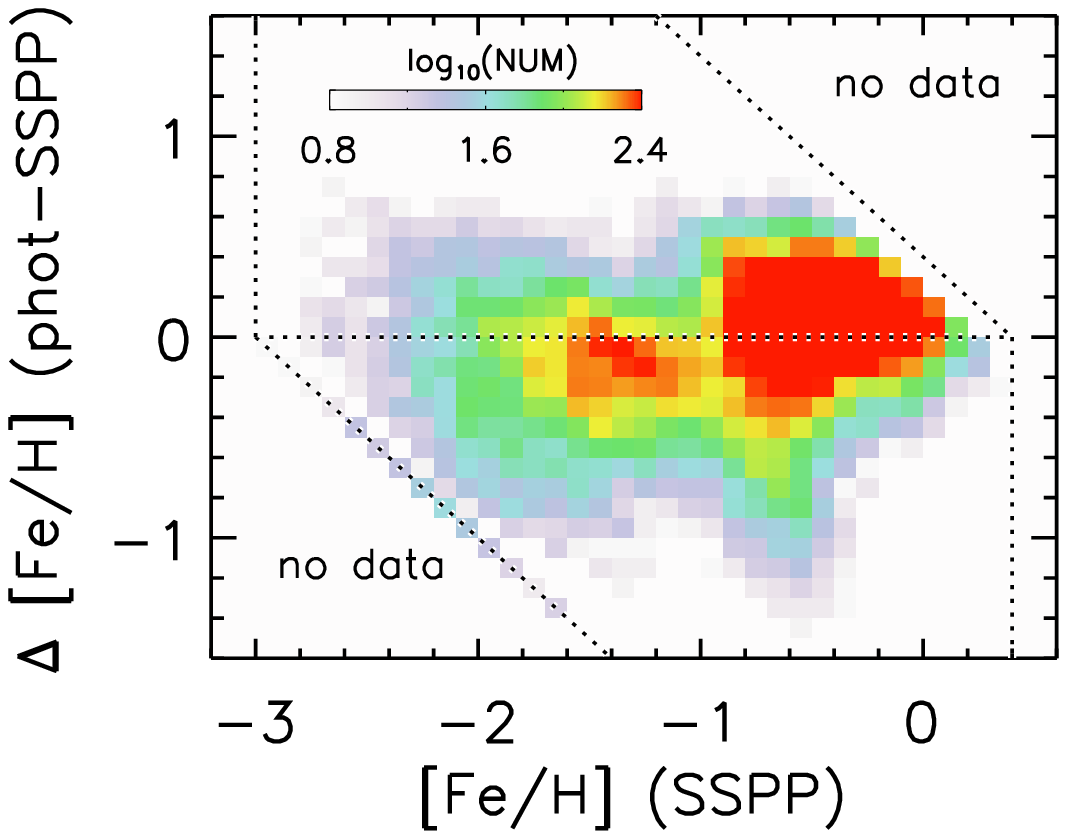}
\caption{Comparison between photometric and spectroscopic (SSPP) metallicity estimates
for the SEGUE sample. The photometric estimates are based on all of the $ugriz$ photometry.
Three curves in the top panel show a median and an interquartile range, respectively.
\label{fig:sspp.comp.feh}}
\end{figure}

Figure~\ref{fig:sspp.comp.feh} shows a comparison between the spectroscopic (SSPP)
metallicities from SDSS/SEGUE and our photometric estimates for the same set
of stars, as in Figure~\ref{fig:sspp.comp.teff}. The central line in the top
panel shows a median photometric metallicity for each $0.2$~dex bin in the
spectroscopic metallicity, and the other two lines represent the first and the
third quartiles. There appears an extended metal-poor tail at a given SSPP [Fe/H]
in Figure~\ref{fig:sspp.comp.feh}, in particular at low [Fe/H] values. This is
because the metallicity sensitivity of stellar colors degrades at lower metal
abundances, due to the non-zero photometric errors. In the next section, we
model these profiles based on simulations of artificial stars. As will be
shown, artificial star tests indicate that the median photometric metallicities
(central line in Figure~\ref{fig:sspp.comp.feh}) are insensitive to photometric
errors and the lower limit in the model grid at [Fe/H]$=-3$, and exhibit a maximum
deviation of only $\Delta {\rm [Fe/H]}\sim0.1$~dex from the true [Fe/H] down to
[Fe/H]$=-2.8$. The overall good agreement between our photometric
metallicities and the SSPP values shown in Figure~\ref{fig:sspp.comp.feh}, with
a maximum deviation of less than $0.2$~dex, confirms that our photometric
metallicity scale is not far from the spectroscopic scale down to [Fe/H]$\sim-2.5$,
perhaps as low as $\sim-3$. However, in order to perform a more stringent test
of the accuracy of our photometric metallicity estimates, we require more
comparison star samples below [Fe/H]$\sim-2.5$.

Because we included $u$-band measurements, which are more sensitive to metal
abundances than the other SDSS bandpasses, the observed dispersion in
Figure~\ref{fig:sspp.comp.feh} is smaller than the equivalent Figure~$2$ in
\citet{an:09b}, which is based only upon $gri$. The $1\sigma$ random scatter
(estimated from the interquartile range divided by $1.349$, assuming
a normal distribution) is about $0.2$--$0.4$~dex for individual stars.
The [Fe/H] estimates in SDSS/SEGUE are precise to $\sim0.2$~dex for individual
stars \citep{lee:08a,lee:08b,allende:08,smolinski:11},
so this comparison indicates that the photometric metallicities can be as
precise as $\sim0.3$~dex per star, when good photometry is available.
The down-turn of the mean trend at [Fe/H]$\sim-0.8$ in Figure~\ref{fig:sspp.comp.feh}
is likely caused by our {\it ad hoc} assumptions on the [$\alpha$/Fe] and age
as a function of [Fe/H] in the model.

An analysis of likely member stars of Galactic open and globular clusters and
high-resolution spectra of SDSS/SEGUE stars indicates that the SSPP exhibits a
tendency to estimate [Fe/H] higher by about $0.25$~dex for stars with
[Fe/H] $<-3.0$, in particular for cool giants. Although the SSPP underestimates
metallicities for stars with super solar abundances, the effect is less than $0.1$~dex.
Nevertheless, it should be kept in mind that the reliability of measuring photometric
metallicities eventually requires a test against high-resolution spectroscopy.

In \citet{an:09b}, we tested the accuracy of photometric metallicities from
the application of $gri$ photometry, and found that the photometric technique
systematically underestimates [Fe/H]$_{\rm phot}$ in the low metallicity range
($\Delta {\rm [Fe/H]}\sim0.3$~dex at [Fe/H]$\sim-1.6$.
We speculated that the mismatch between the SSPP and photometric metallicities
were likely caused by either unresolved binaries in the sample or a small
zero-point offset in the model colors. Although the effect of unresolved
binaries in the photometric metallicities cannot be completely ignored (see
\S~\ref{sec:artificial} below), small systematic color differences can actually
induce a systematic trend in [Fe/H]$_{\rm phot}$, especially when photometric
metallicities are derived based on color indices that are weakly dependent on
[Fe/H], such as the application of $gri$ photometry in \citet{an:09b}.
The accuracy of the photometric technique (Figure~\ref{fig:sspp.comp.feh})
has improved as more metallicity-sensitive color indices ($ugriz$) are employed
in the calculation of [Fe/H]$_{\rm phot}$.

\begin{figure}
\centering
\includegraphics[scale=0.65]{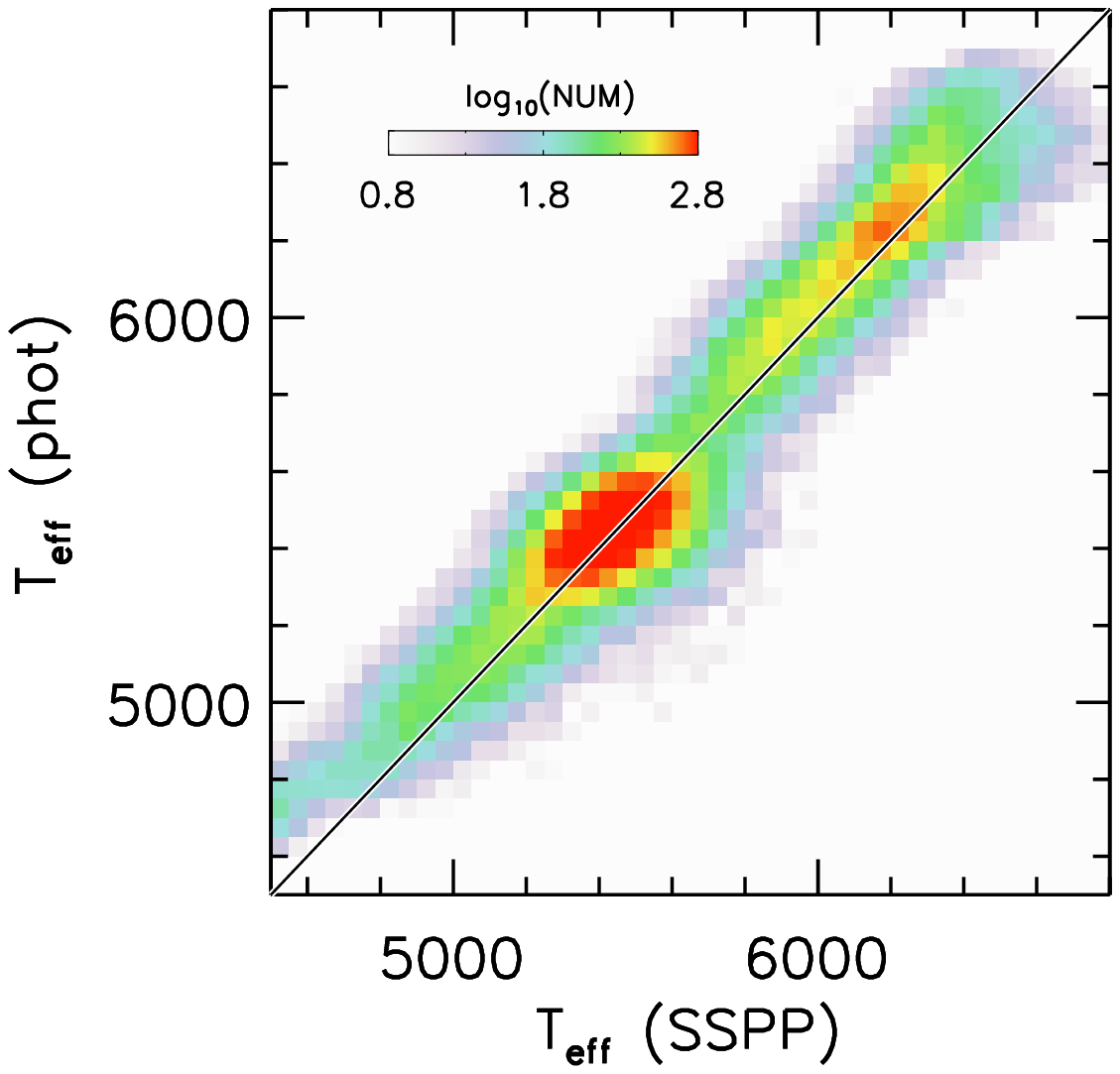}
\includegraphics[scale=0.65]{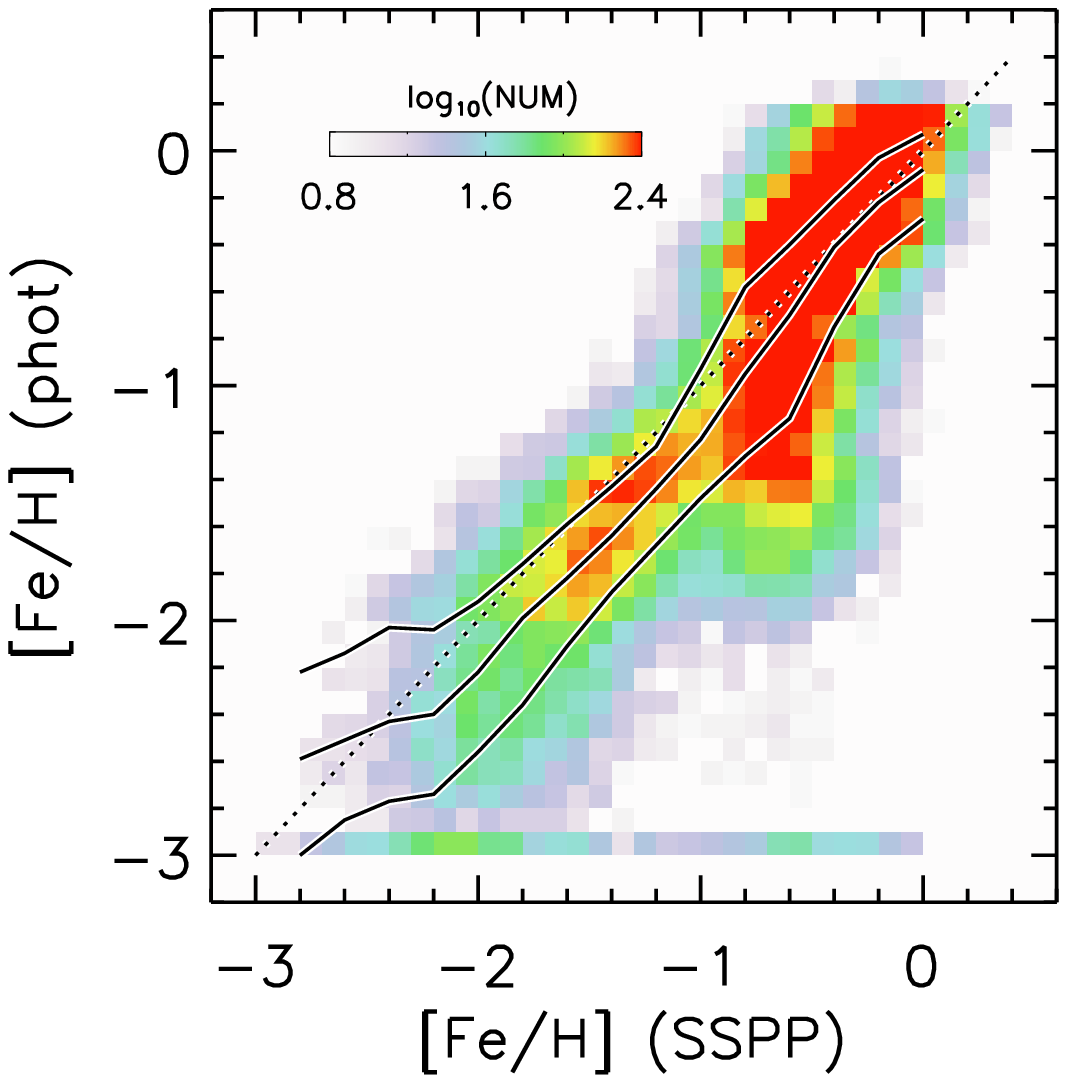}
\caption{Comparisons between photometric and spectroscopic (SSPP) temperature
({\it top panel}) and metallicity estimates ({\it bottom panel}) for the SEGUE sample.
Photometric estimates are based only on $ugr$ photometry. Three curves in the bottom
panel show a median and an interquartile range, respectively.
\label{fig:sspp.comp.ugr}}
\end{figure}

Figure~\ref{fig:sspp.comp.ugr} is the same as
Figures~\ref{fig:sspp.comp.teff}--\ref{fig:sspp.comp.feh},
but based only on $ugr$ colors. The scatter in the comparison is larger than those
seen in Figures~\ref{fig:sspp.comp.teff}--\ref{fig:sspp.comp.feh}, because of the
weaker constraints on these parameters. The down-turn of the mean trend at
[Fe/H]$\sim-0.8$ is stronger than seen in Figure~\ref{fig:sspp.comp.feh}. Clearly,
the inclusion of all $ugriz$ passbands is preferred, in order to strongly constrain
both photometric temperatures and metal abundances.

\subsection{Artificial Star Tests}\label{sec:artificial}

We performed artificial star tests in order to evaluate the effects of
photometric errors and unresolved binaries and/or blends on our determinations
of photometric [Fe/H] estimates. In particular, we use these simulation results
to construct metallicity kernels, which are employed to assess the broadening of the
[Fe/H] profiles resulting from the above effects. These kernels are then used in
\S~\ref{sec:results} to deconvolve the observed MDF of the halo.

\begin{figure*}
\epsscale{0.95}
\plotone{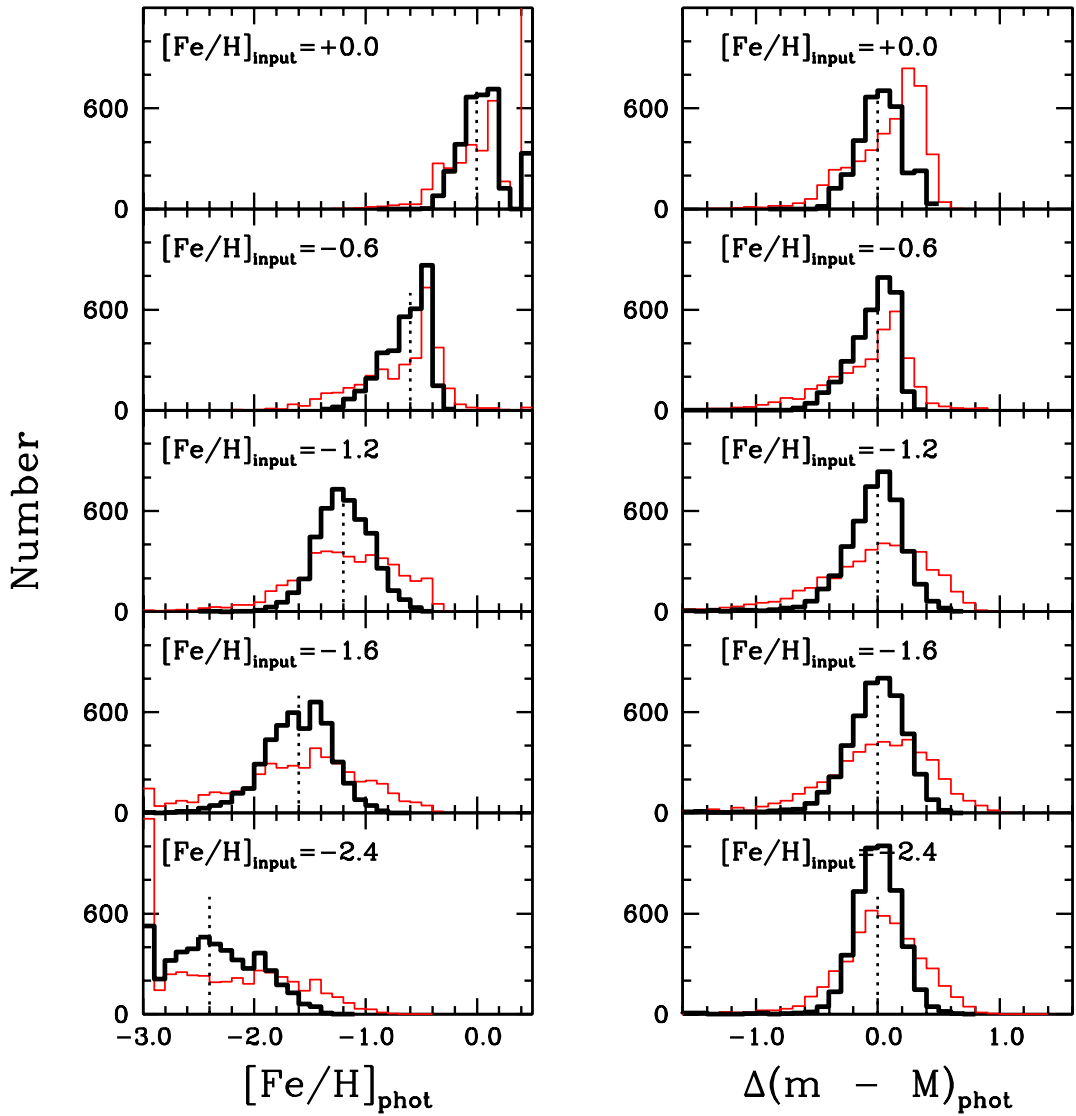}
\caption{Artificial star tests for various metallicity bins. {\it Left:} The thick black
histogram shows a photometric MDF, where photometric errors of $0.02$~mag in
$gri$ and $0.03$~mag in $u$ and $z$ passbands were assumed. The thin red
histograms show the case when photometric errors are twice the size of these
errors. Input [Fe/H]$_{\rm input}$ values are shown on each panel, and indicated
by the dotted vertical lines. No binaries are included in the simulation.
{\it Right:} Same as in the left panels, but for the displacement in distance modulus.
\label{fig:single}}
\end{figure*}

Figure~\ref{fig:single} shows the artificial star test results for a single
stellar population without binaries. In this  exercise, we used the same
set of models as in \S~\ref{sec:calibration}, which include the main sequence
and a portion of the subgiant branch.
For each of the various metallicity bins,
we generated 20,000 artificial stars with masses above $0.65 M_\odot$ from
a model isochrone. The magnitudes were then convolved with a Gaussian having
a standard deviation of $0.02$~mag in $gri$, and $0.03$~mag in the $u$ and $z$
passbands. We derived photometric metallicities and distance estimates as described
in \S~\ref{sec:param}; the resulting MDFs are shown as solid black histograms
in the left panels, and the distributions in distance modulus are in the right panels.
We used the following selection criteria:
\begin{itemize}
  \item $\chi^2_{\rm min}/\nu < 3.0$, where $\chi^2_{\rm min}$ is a minimum
        $\chi^2$ defined in Equation~(\ref{eq:chi}) and $\nu$ ($=2$) is a degree of freedom
  \item $\log{g}_{\rm (YREC)} \geq 4.15$
  \item $0.65 < M_*/M_{\rm sun} < 0.75$.
\end{itemize}
The last criterion is specified in order to be consistent with the analysis
of the actual data sets described in \S~\ref{sec:results}, and assumes that stars
are on the main sequence.

The solid red histograms shown in Figure~\ref{fig:single} are those resulting
from the simulations, but assuming photometric errors twice the size as above
($0.04$~mag error in $gri$, $0.06$~mag error in $u$ and $z$, and assuming no
correlation between the errors in different bandpasses), illustrating the effect
of the size of photometric errors on the photometric metallicity estimates. For
a given photometric error, the resulting dispersion is higher at lower
metallicity, due to the weaker dependence of broadband photometric colors on
metallicity at lower abundances; e.g., see Figure~$1$ in \citet{an:09b}. At
solar abundance, the standard deviation of the [Fe/H] distribution in
Figure~\ref{fig:single} is $0.1$~dex, but increases to $0.3$~dex at
[Fe/H]$=-1.6$, and $0.5$~dex at [Fe/H]$=-2.4$. At the lowest [Fe/H], the
distribution becomes asymmetric, with an extended low-metallicity tail.
On the other hand, the distribution in distance modulus becomes progressively
more symmetric at the lower input metallicity, because of the weaker metallicity
sensitivity of colors and magnitudes at lower metal abundances.
Our model set does not extend below [Fe/H]$=-3$ or above [Fe/H]$=+0.4$,
which results in a piling up of stars at these metallicities.

\begin{figure*}
\epsscale{1.0}
\plotone{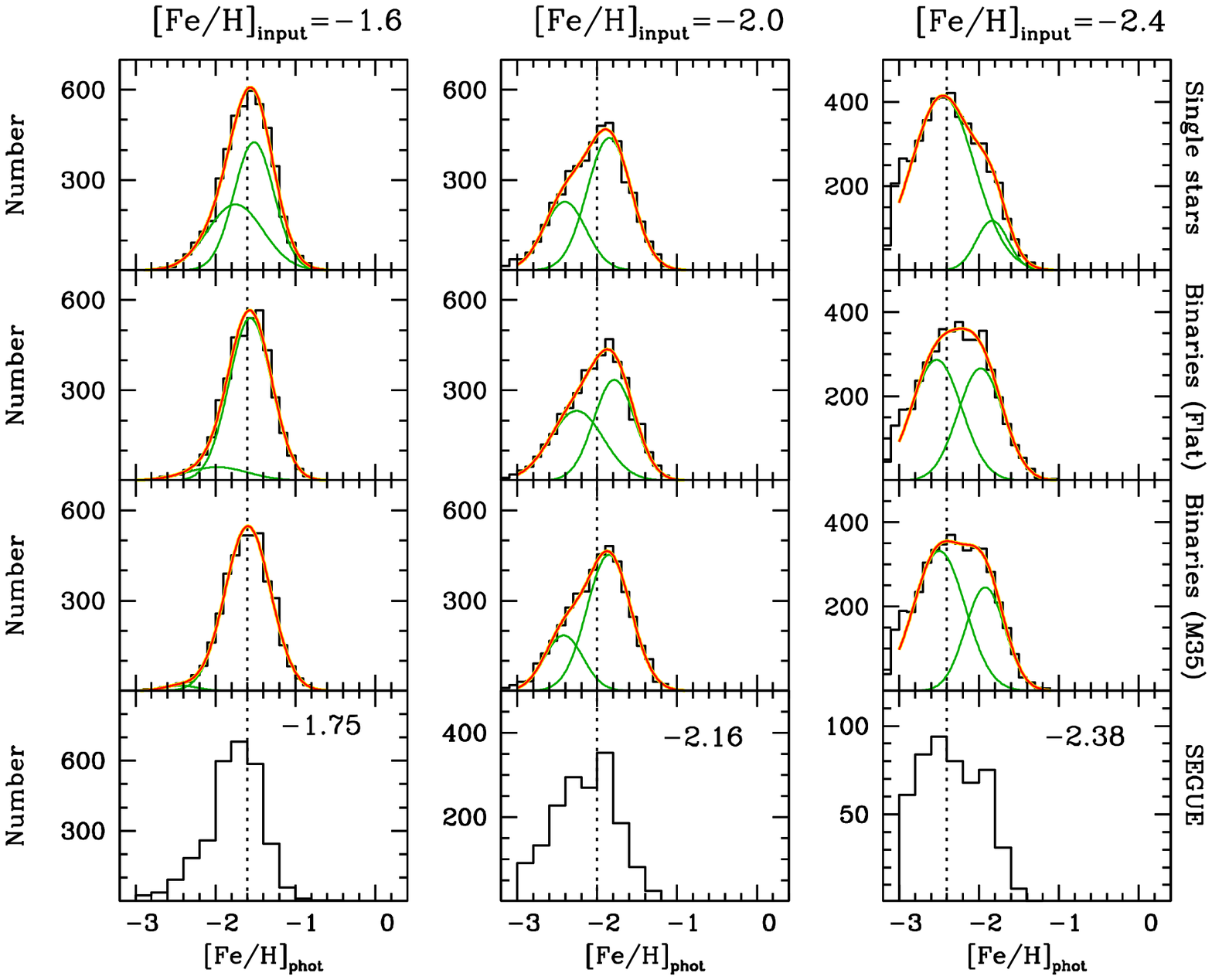}
\caption{Effects of unresolved binaries and/or blends from artificial star tests
at three input metallicity bins: [Fe/H]$_{\rm input}=-1.6$ (left), $-2.0$ (middle), and
$-2.4$ (right). {\it Top:} A single star population without binaries. {\it Middle:}
A 50\% unresolved binary fraction with a flat, and a M35-type mass function, respectively,
for secondaries. The green curves in each panel show Gaussian fits to the distribution;
the red curve is their sum. {\it Bottom:} Photometric metallicity distributions of
SDSS/SEGUE spectroscopic samples in the same metallicity ranges:
$-1.7 \leq {\rm [Fe/H]_{SSPP}} < -1.5$ (left), $-2.1 \leq {\rm [Fe/H]_{SSPP}} < -1.9$
(middle), $-2.5 \leq {\rm [Fe/H]_{SSPP}} < -2.3$ (right).
Median [Fe/H]$_{\rm phot}$ values are shown in the panels.
\label{fig:binary}}
\end{figure*}

Figure~\ref{fig:binary} shows the effects of unresolved binaries and/or blends
for three metallicity bins, [Fe/H]$_{\rm input}=-1.6$ (left panels), $-2.0$
(middle panels), and $-2.4$ (right panels), respectively. Photometric errors of
$0.02$~mag in $gri$ and $0.03$~mag in $u$ and $z$ passbands were assumed. To
reduce the effects of interpolation errors in the isochrones, which sometimes
produce a non-continuous distribution in [Fe/H]$_{\rm phot}$, we convolved the
derived photometric metallicities with Gaussians having the grid size of the model
($\sigma_{\rm [Fe/H]}=0.1$~dex). The top panels apply to stellar populations
comprised entirely of single stars. The middle two panels show the results when
unresolved binaries are included in the sample. In total, 10,000 primary stars
were generated from a single [Fe/H] population, and the same number of secondary
stars was generated, based either on a flat mass function, or adopting
the M35 mass function \citep{barrado:01}. These simulated stars were merged
with 10,000 single stars with no secondary components to simulate a 50\%
unresolved binary fraction. The resulting magnitudes in $ugriz$ were convolved
with Gaussians having the specified photometric errors.

As can be appreciated by inspection of Figure~\ref{fig:binary}, the presence of
unresolved binaries has little impact on the derived photometric [Fe/H] distribution,
because the dispersion is dominated by the effects of photometric errors.
Binaries, due to the modified flux distributions due to a secondary component,
typically have lower photometric metal
abundances than those of the primaries alone, but these effects are mostly buried
in the extended [Fe/H] distribution produced by the presence of non-zero photometric errors.
The bottom panels in Figure~\ref{fig:binary} show photometric metallicity
distributions of the SDSS/SEGUE spectroscopic sample in
Figures~\ref{fig:sspp.comp.teff}--\ref{fig:sspp.comp.feh} in the same metallicity
bins: $-1.7 \leq {\rm [Fe/H]_{SSPP}} < -1.5$ (bottom left),
$-2.1 \leq {\rm [Fe/H]_{SSPP}} < -1.9$ (bottom middle),
$-2.5 \leq {\rm [Fe/H]_{SSPP}} < -2.3$ (bottom right). Except for a small offset
in the metallicity scale between the SSPP and photometric metallicities
(median [Fe/H]$_{\rm phot}$ values are shown in the panels), which has
already been pointed out in \S~\ref{sec:sspp}, the overall [Fe/H]$_{\rm phot}$
distributions appear quite similar to those from the artificial star tests.

Our strategy in this study is to deconvolve the observed MDF of halo stars using
simulated profiles, and infer the underlying [Fe/H] distribution
(\S~\ref{sec:results}). In order to carry out the deconvolution in a
straightforward manner, we parameterized simulated [Fe/H] kernels by
simultaneously fitting two Gaussian functions, as shown by the two green curves
in each panel of Figure~\ref{fig:binary}. The red curve is a sum of these two
distributions. As discussed above, the precise binary fractions and the exact
form of the adopted mass functions are not dominant factors in determining the
overall shape of a given MDF. Therefore, we opted to choose the M35 mass
function with a 50\% binary and/or blending fraction when comparing the observed
MDF of the halo with simulated results.

\begin{figure}
\epsscale{1.15}
\plotone{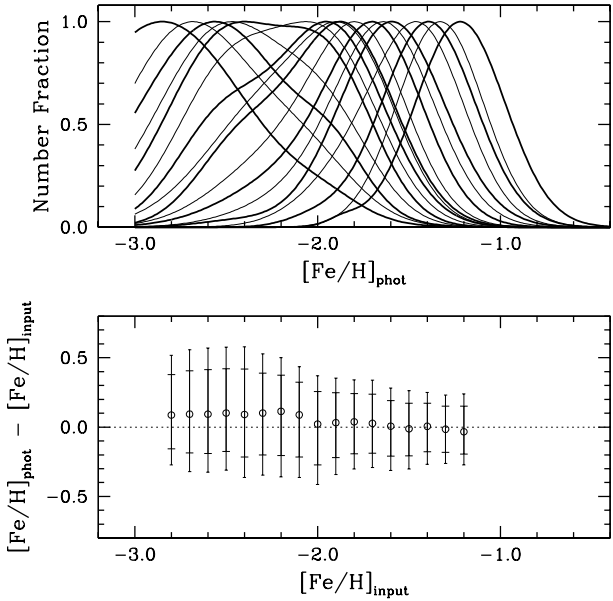}
\caption{{\it Top:} Normalized distributions of photometric metallicities at
various input [Fe/H]$_{\rm input}$ values. These simulated profiles were generated
from artificial star tests at a given input [Fe/H]$_{\rm input}$, assuming a 50\%
binary fraction and a M35-type mass function, and by taking $0.02$--$0.03$~mag
errors in $ugriz$ (see text). Input [Fe/H] values are from $-2.8$ to $-1.2$ in
an $0.1$~dex increment, where simulated profiles are shown alternatively in
thick and thin curves. {\it Bottom:} Median differences between photometric
metallicities and input values. Longer error bars show interquartile ranges,
and the shorter error bars represent $\pm1\sigma$ errors. The latter values were
computed from the interquartile ranges divided by $1.349$.
\label{fig:fehprofile}}
\end{figure}

The top panel in Figure~\ref{fig:fehprofile} displays photometric metallicity kernels for various
input metallicity values ([Fe/H]$_{\rm input}$), normalized to the peak value of
each profile. These simulated profiles were generated from the same set of
artificial star tests as above, assuming a M35-type mass function and by taking
$0.02$~mag error in $gri$ and $0.03$~mag error in $u$ and $z$. The input
[Fe/H]$_{\rm input}$ values are from $-1.2$ to $-2.8$ in intervals of $0.1$~dex;
for clarity, the simulated profiles are alternatively shown in thick and thin
curves. Again, note that the resulting photometric distribution becomes broader
at lower metallicities.

The bottom panel in Figure~\ref{fig:fehprofile} shows the median differences between
photometric metallicities and input [Fe/H] values for the same data set used to
construct the kernels in the top panel. Longer error bars represent the interquartile
ranges, while the shorter error bars represent $\pm1\sigma$ errors, where we computed the
$1\sigma$ error by dividing the interquartile ranges by $1.349$, assuming a normal
distribution.

\section{Sample Selection and Properties}\label{sec:sample}

Using the new photometric metallicity estimation technique described in the
previous section, we estimated distances and metal abundances for individual
stars in SDSS Stripe~82, which is one of the imaging stripes in SDSS that has
been repeatedly scanned along the celestial equator. There are two photometric
catalogs available in Stripe~82: the {\it calibration} or Standard Star catalog
\citep[][hereafter calibration catalog]{ivezic:07}\footnote{Available at
{\tt http://www.astro.washington.edu/\\users/ivezic/sdss/catalogs/stripe82.html}.}
and the {\it coadd} imaging
catalog \citep[][hereafter coadded catalog]{annis:11}. The calibration star catalog
contains stellar magnitudes for approximately one million sources, where the
magnitudes were averaged at the catalog level. The coadded catalog is based on
the coadded image products, and is about $0.5$~mag deeper than the calibration
catalog. Both catalogs, which were in principle produced from the same
observations, obtained with the ARC $2.5$-m SDSS survey telescope facilities,
provide the most precise ($\sim1\%$) photometry set available within SDSS, and
therefore can be used to set the best available constraints on the photometric
MDF of the Galaxy.
Both catalogs were constructed on the {\tt Photo} magnitude system. However,
we directly employed our {\tt UberCal}-based models in the parameter estimation
using this photometry, because the global photometric zero-point differences
between the two systems are negligible.

Below, we describe the selection of a photometric sample from Stripe~82 designed
to minimize bias (\S~\ref{sec:bias}), and evaluate the effects of unrecognized
giants in the sample (\S~\ref{sec:giants}). In the subsequent section
(\S~\ref{sec:results}), we present an unbiased MDF of the Galactic halo, and test
the hypothesis that the halo is composed of two overlapping sub-components,
based on both metallicity and kinematic information for our sample stars.

\subsection{Sample Selection and Bias}\label{sec:bias}

An unbiased sample of stars is, of course, an important ingredient for obtaining
a representative MDF of the Milky Way's halo population(s). Although photometric
samples are less susceptible to sample biases than spectroscopic studies
that make use of metallicity or color in their sample selection,
there still exists a bias that needs to be taken into account, as discussed below.

\begin{figure}
\epsscale{1.15}
\plotone{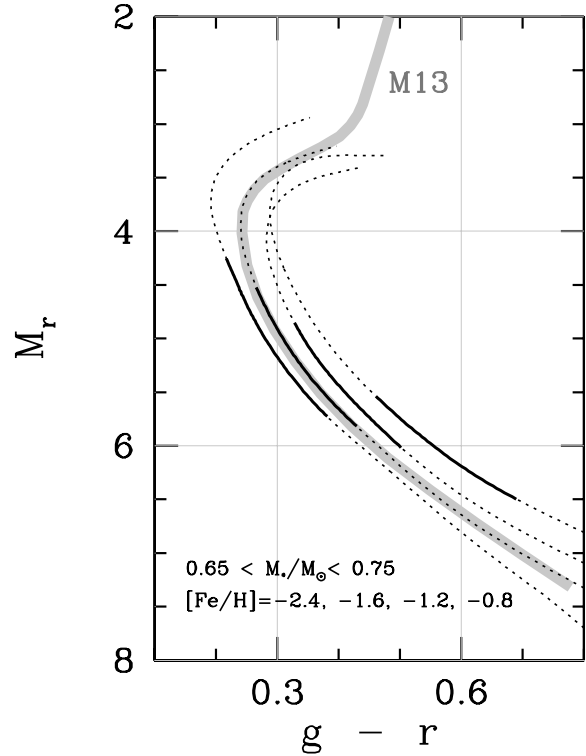}
\caption{The range of stellar mass in the model considered in this work.
Color-calibrated models
are shown at [Fe/H]$=-2.4$ (13~Gyr; left-most), $-1.6$ (13~Gyr), $-1.2$ (13~Gyr), and
$-0.8$ (9~Gyr; right-most), where soild lines represent $0.65 < M_*/M_\odot < 0.75$.
The fiducial sequence of M13 is shown as a gray line for comparison.
\label{fig:cmd}}
\end{figure}

In this work we adopted a sample selection based on stellar mass, as estimated
using our isochrones. Figure~\ref{fig:cmd} shows our color-calibrated models at
[Fe/H]$=-2.4$, $-1.6$, $-1.2$, and $-0.8$; the thick solid lines indicate where
$0.65 < M_*/M_\odot < 0.75$. This range of stellar mass is similar to what is adopted
in this work (see below). Our choice for the mass-based sample selection was motivated
by our reasoning that, in order to obtain a representative sample of the halo, stars
at different metallicities should be sampled in identical mass ranges.
We consider this to be a superior choice to the more commonly adopted color-based
selections, because of the strong relationship between color, metallicity,
and mass; narrow color-cuts in a stellar sample would produce a mix of stars,
including less massive, lower metallicity stars and more massive, higher
metallicity stars.  Although colors can be used as a surrogate for temperatures,
they have only a limited applicability for stellar masses, hence we believe our
mass-based selection should produce a less biased sample of stars for the
assembly of a valid halo MDF.

\begin{figure}
\epsscale{1.15}
\plotone{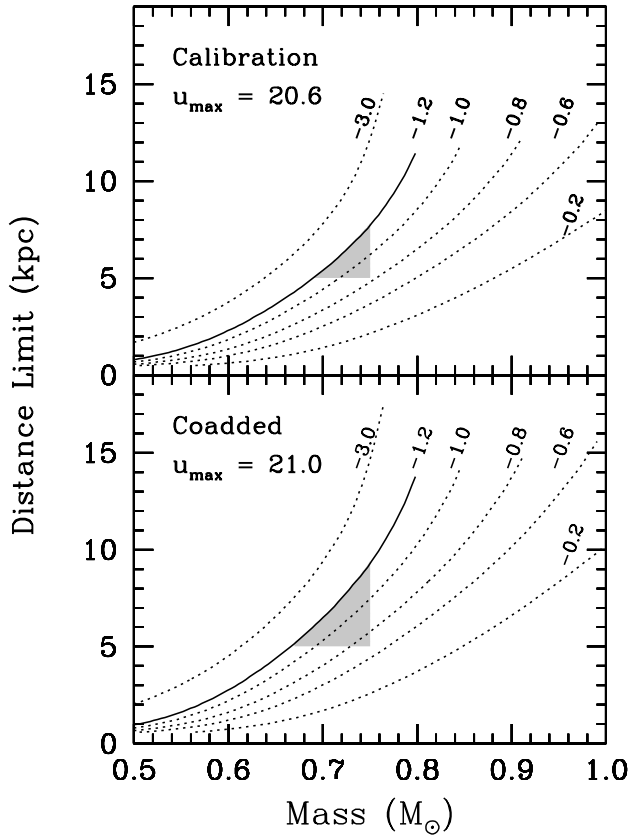}
\caption{Maximum heliocentric distance as a function of stellar mass
that can be reached by a star at $u < 20.6$~mag (top panel) and $u < 21.0$~mag
(bottom panel). These magnitude limits correspond to $\sigma_u \approx 0.03$~mag error in $u$ in
the Stripe~82 calibration (top) and coadded (bottom) catalogs, respectively. The solid and
dotted curves show distance limits at a number of different metallicity bins. The gray shaded
region represents a mass-distance limit set for the halo sample in this work, which ensures
that the sample is unbiased at ${\rm [Fe/H]} \la -1.2$, has a stellar mass less than the
turn-off mass of the [Fe/H]$=-3$ model, and is relatively free from thick-disk contamination
($d_{\rm helio} > 5$~kpc).
\label{fig:maxdist}}
\end{figure}

Figure~\ref{fig:maxdist} illustrates metallicity-luminosity relations as a
function of stellar mass, where the luminosity is expressed in terms of a
maximum heliocentric distance that can be reached by a star at a specific
magnitude limit.
Note that we adopted an age of $13$~Gyr for models at
$-3.0 \leq {\rm [Fe/H]} \leq -1.2$, and $4$~Gyr at $-0.3 \leq {\rm [Fe/H]} \leq +0.4$,
with a linear interpolation between these two metallicity ranges (\S~\ref{sec:isochrone}).
The $u$-band magnitude limit was used in
Figure~\ref{fig:maxdist}, because of the strong sensitivity of this band on
metal abundances. The $u_{\rm max} = 20.6$ (top panel) and $u_{\rm max} = 21.0$
(bottom panel) correspond to a median photometry error of $\sigma \approx 0.03$
mag in the calibration and the coadded catalogs, respectively. The error size is
similar to those adopted in the artificial star tests (\S~\ref{sec:artificial}).

At a given $u$-band magnitude limit, we computed a maximum distance to which
each star can be observed at a given stellar mass and metal abundance. At a
fixed mass, metal-poor stars are brighter than metal-rich stars, thus they can
be observed at greater distances than metal-rich stars (see Figure~\ref{fig:cmd}).
Note that a color-based
selection would have the opposite consequence on the luminosity of stars that
would be included; at a fixed color (or temperature) stars are brighter at
higher metallicity. It follows that, in a magnitude-limited survey such as SDSS,
color-based selection would produce samples that are biased against more
metal-poor stars at greater distances from the Sun.

\begin{figure}
\epsscale{1.15}
\plotone{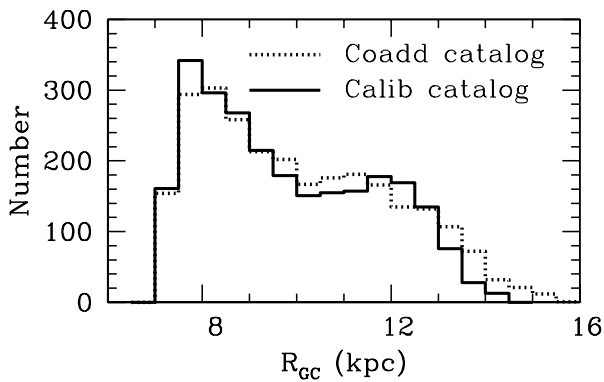}
\caption{Distribution of Galactocentric distances for the final halo samples in
the calibration and the coadded catalogs in Stripe~82.
\label{fig:gcdist}}
\end{figure}

The gray shaded region in Figure~\ref{fig:maxdist} indicates the mass-distance
limit set in our halo sample. In both panels, these areas are surrounded by a
[Fe/H]$=-1.2$ model to insure that the sample is unbiased at ${\rm [Fe/H]} \leq
-1.2$. Our photometric halo MDFs should be less affected by this choice, since
we generally expect to obtain relatively few halo stars outside this metallicity
range. An additional constraint on the stellar mass has been applied to the
sample, because main-sequence turn-off masses (the high-mass end points of each
curve in Figure~\ref{fig:maxdist}) are varying at different metallicities.
Therefore, we imposed an upper mass limit, $M_* < 0.75 M_\odot$, which is close
to the main-sequence turn-off mass from a [Fe/H]$=-3$ model; turn-off masses
occur at higher masses for [Fe/H]$>-3$. The lower limit to the heliocentric
distance ($5$~kpc) was set to exclude possible thick-disk interlopers in the
sample; this naturally results in a lower mass cut-off at $\sim0.65\ M_\odot$.
We further imposed a cut based on Galactic latitude (see discussion below
in \S~\ref{sec:mdf}). The application of these constraints results in a
relatively narrow parameter space (gray areas in Figure~\ref{fig:maxdist}) in
the stellar mass versus distance plane ($5$~kpc $\leq d_{\rm helio} \la 8$~kpc
for the calibration catalog; $5$~kpc $\leq d_{\rm helio} \la 9$~kpc for the
coadded catalog). In short, our sample selection (delineated by the gray
regions in Figure~\ref{fig:maxdist}) collects all of the stars at [Fe/H]$\la-1.2$
in a limited volume of the halo, given the magnitude limit set in a photometric
catalog ($\sigma_u < 0.03$~mag in this work).
The resulting distribution of our sample in Galactocentric distances is shown in
Figure~\ref{fig:gcdist}.

We have carried out comparisons of our samples obtained by both the calibration
and the coadded catalogs, using a search radius of $1\arcsec$. As recommended in
\citet{ivezic:07}, we only included sources in the calibration catalog with
at least four repeated observations in each bandpass, for both the photometric
comparisons and the following analysis, and proceed with the mean magnitudes and
their standard errors. After carrying out iterative $3\sigma$ rejections, we
found zero-point differences of $-0.007\pm0.022$ mag, $-0.001\pm0.011$ mag,
$+0.002\pm0.013$ mag, and $+0.003\pm0.015$ mag, in $u\, -\, g$, $g\, -\, r$,
$g\, -\, i$, and $g\, -\, z$, respectively, for stars with $u < 20.6$ and
fainter than $16^{\rm th}$ magnitude in each filter bandpass;
see also photometric comparisons in \citet{annis:11}. The sense of these
differences is that the $u\, -\, g$ color measurements in the coadded catalog
are redder than those in the calibration catalog. We restricted our comparison
to only include those stars with Galactic latitudes at $|b| > 35\arcdeg$, to
be consistent with the analysis of the halo MDF carried out below. The listed
errors indicate the derived dispersions in color difference (rather than the
errors in the mean), using $\sim100,000$ objects in each comparison. Because
fainter $u$-band measurements yield less strong UV excesses, the use of
photometry from the coadded catalog would lead to systematically higher
photometric metallicity estimates than in the calibration catalog (this is
confirmed in the next section). The rms deviations of the color differences
between the two catalogs are stable ($\sigma_{\rm color} < 0.003$~mag) along the
$110\arcdeg$ length of Stripe~82.

Although a detailed study is beyond the scope of the present work,
we identified a strong systematic deviation in the $u$ passband
($\Delta u \sim0.2$~mag at $u\sim22$~mag) at the faint end, beyond the
magnitude limit set in our sample ($u < 20.6$).
One likely cause of the systematic offset is a Malmquist-type bias in the
calibration catalog, because faint sources near the detection limit can
either be detected or missed, due to the existence of large Poisson errors.
Since the calibration catalog took the average of the source magnitudes from
individual images, the mean magnitude could therefore be biased towards
brighter source measurements. We avoided the photometry zero-point issue by
requiring $\sigma_u < 0.03$ mag; small photometric errors also make the
photometric metallicity estimates more reliable.

We have not made use of the main SDSS photometric database in this work. The $95\%$
completeness limit in the $u$ bandpass of the main survey is $22.0$~mag, but the
$\sigma_u = 0.03$ mag limit corresponds to $u \approx 18.7$~mag, on the order of
$2$~mag shallower than for the Stripe~82 catalog. At this limiting magnitude,
use of the main SDSS database would allow exploration of heliocentric distances
only up to $\sim2.2$~kpc for the construction of an unbiased [Fe/H] sample at
[Fe/H]$< -1.2$, which is an insufficient volume to probe the halo MDF.

To summarize, in the following analysis we used the selection criteria listed below:
\begin{itemize}
  \item Sources are detected in all five bandpasses
  \item $\chi^2_{\rm min}/\nu < 3.0$, where $\chi^2_{\rm min}$ is a minimum
        $\chi^2$ defined in Equation~(\ref{eq:chi}) and $\nu$ ($=2$) is a degree of freedom
  \item $\log{g}_{\rm (YREC)} \geq 4.15$
  \item $0.65 M_\odot < M_* < 0.75 M_\odot$
  \item Mass-metallicity-luminosity limits (gray region in Figure~\ref{fig:maxdist})
  \item $d_{\rm helio} > 5$~kpc
  \item $|b| > 35\arcdeg$.
\end{itemize}

The above selection restricts the photometric sample to $0.2 \la g\, -\, r \la 0.5$
(see Figure~\ref{fig:cmd}). The total numbers of objects that passed the above selection
criteria are $2,523$ from the calibration catalog and $2,626$ from the coadded catalog.
The median $r$ magnitudes are $18.7$~mag and $19.1$~mag for the calibration and the coadded
catalogs, respectively, with a $\sim0.3$~mag dispersion. The standard SDSS star-galaxy
separation based on the difference between PSF (Point Spread Function) and model magnitudes
is robust to $r\sim21.5$ \citep{lupton:02,scranton:02}, so the contamination by galaxies
in our sample should be negligible \citep[see also][]{annis:11,bovy:12a}.

\subsection{Contamination by Distant Giants and Thick Disk Stars}\label{sec:giants}

Our calibration is valid for main-sequence stars only, and we have explicitly
assumed that the great majority of stars in the sample are in their
main-sequence phase of evolution. However, there certainly exist distant
background giants and subgiants along each line of sight, and their distances can be greatly
underestimated if a photometric parallax relation for main-sequence stars is
blindly applied. Unfortunately, $ugriz$ photometry alone is not sufficient
to reliably discriminate giants from dwarfs, unlike horizontal-branch stars
\citep{ivezic:07}, and therefore some level of contamination by giants in our
sample is unavoidable.

We evaluated the level of expected giant and subgiant contamination using
cluster photometry for both M13 ([Fe/H]$ =-1.6$; see Figure~\ref{fig:cmd})
and M92 ([Fe/H] = $-2.4$) \citep{an:08}. In order to carry this out, we applied
cuts based the on $\chi$ and {\tt sharp} parameters in DAOPHOT \citep{stetson:87}
as in \citet{an:08}. Based on the large extent of these clusters on the SDSS CCD
chips, we opted not to apply cuts based on the distance from the cluster centers.

We assumed that the number density of the halo follows a power-law profile with
an index of $-2.8$, and a halo ellipticity of $0.6$ \citep[e.g.,][]{juric:08}.
Using this model, along with the photometry for each cluster, we simulated
color-magnitude diagrams for each line of sight in Stripe~82, and applied the
same mass-metallicity-luminosity cuts to the sample as in the previous section.
We integrated the number of giants out to $30$~kpc from the Sun, beyond which
the giant contamination is negligible, because either those distant giants are
too faint to be included in our sample or they do not satisfy our selection
criteria. Each object was tagged as either a (sub)giant or a dwarf, depending on the
location on the original cluster color-magnitude diagram, and the number of
giants included in the sample was counted. From the above calculation we found
that the contamination rates are at about the $10\%$--$15\%$ level from the M13
photometry, and the $15\%$--$20\%$ level when using the M92 photometry. M92 is
more metal-poor than M13, so more cluster giants fall in the color range ($0.2
\la g\, -\, r \la 0.5$), which is implicitly set by our sample selection
criteria. This results in a slightly higher contamination rate from the M92
photometry.

More than any other parameters (e.g., shape parameters of the halo) in the model,
we found that the assumed fraction of giants is the dominating factor that determines
the total contamination rate in our sample. Note that our estimate for giant
contamination in the sample is higher than what \citet{juric:08} estimated
($\sim15\%$ vs. $\sim4\%$). \citet{juric:08} obtained an estimate of the fraction
of giants to be about $5\%$ from M13, using the SDSS pipeline {\tt Photo} values
in this crowded region, over a color range similar to that in our analysis, and
concluded that the bias in the number density of the halo is about $4\%$ arising
from the misidentification of giants as main-sequence dwarfs.

\begin{figure*}
\epsscale{0.95}
\plotone{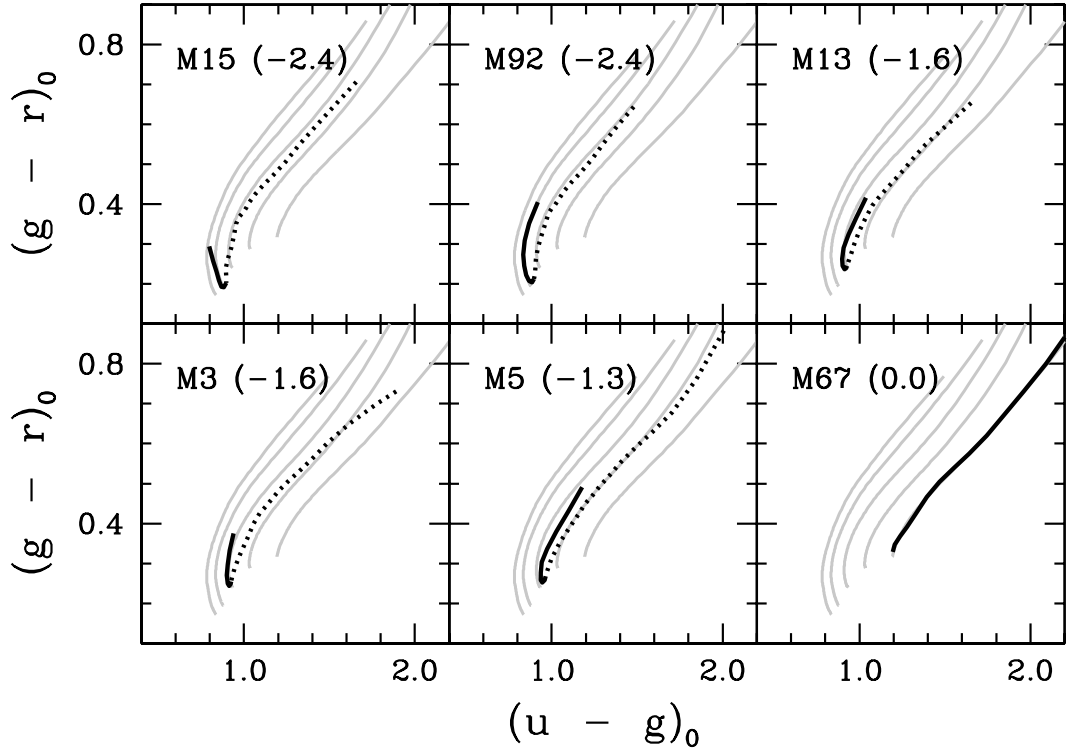}
\caption{Fiducial sequences of the calibration cluster sample (black lines) in
the $u\, -\, g$ vs.\ $g\, -\, r$ plane. The solid black lines represent a dwarf
sequence; dotted lines show the giant sequence. Overlaid gray lines are calibrated models
at [Fe/H]$=-3.0$ (left), $-2.4$, $-1.6$, $-0.8$, and $0.0$ (right), respectively.
At a given $g\, -\, r$, giants have redder $u\, -\, g$ colors than the main sequence,
leading to an overestimated photometric metallicity. Values in parenthesis next to
the cluster name indicate the cluster metal abundance in [Fe/H].
\label{fig:ugr}}
\end{figure*}

Giant contamination in our sample produces an overall shift of the
photometric metallicity estimates toward higher values. Figure~\ref{fig:ugr} shows the
fiducial sequences on the $u\, -\, g$ versus $g\, -\, r$ diagram for a number of
clusters that were used in our color calibration. The black lines shown are fiducial
sequences, with the dotted lines indicating a giant sequence.  For M67, only a 
main sequence was detected in SDSS \citep[giants are too bright to be included;][]{an:08}.
The overlaid gray lines in each panel show our calibrated models at
[Fe/H]$=-3.0$ (left most), $-2.4$, $-1.6$, $-0.8$, and $0.0$ (right most),
respectively. At a given $g\, -\, r$, giants have redder $u\, -\, g$ colors than
the main sequence, which leads to an overestimated photometric metallicity if
they are misidentified as dwarfs. The size of a bias in the photometric
metallicity can be as large as $0.5$--$1$~dex, but fortunately this only applies
for a limited number of stars along each line of sight.

On the other hand, contamination by thick-disk stars in our sample is negligible.
We performed a set of Galactic simulations to check the overall fraction of
thick-disk interlopers in our sample along various lines of sight, using artificial
star test results in \S~\ref{sec:artificial}. To simulate a dispersion in the
underlying [Fe/H] distribution, we combined artificial stars at the central [Fe/H]
($\approx-0.7$ for the thick disk and $\approx-1.6$ for the halo) with
those at $\pm1\sigma$ values ($\pm0.2$ and $\pm0.4$~dex for the thick disk and
the halo, respectively), taking normalizations from a Gaussian distribution.
We adopted the Galactic structural parameters in \citet{juric:08}.

The fraction of thick-disk stars, of course, is varying at different Galactic
latitudes, so we computed the fraction of thick-disk stars along Stripe~82.
We found that the average fraction is negligible ($0.4\%$ of the entire sample)
below photometric [Fe/H]$=-1.0$ \citep[see also][]{bovy:12b}. This is because
our sample selection, based on mass, metallicity, and distance, is strongly
biased against stars with [Fe/H]$>-1.2$. If all the stars below solar metallicity
are included in this estimate, the fraction becomes $\sim3.4\%$. However, only
stars with photometric metallicities less than [Fe/H]$=-1$ are concerned in the
following discussions, and we can safely assume that the thick-disk contamination is
negligible.

\section{Results}\label{sec:results}

\subsection{The Observed Metallicity Distribution Function of the Galactic Halo}\label{sec:mdf}

\begin{figure}
\epsscale{1.15}
\plotone{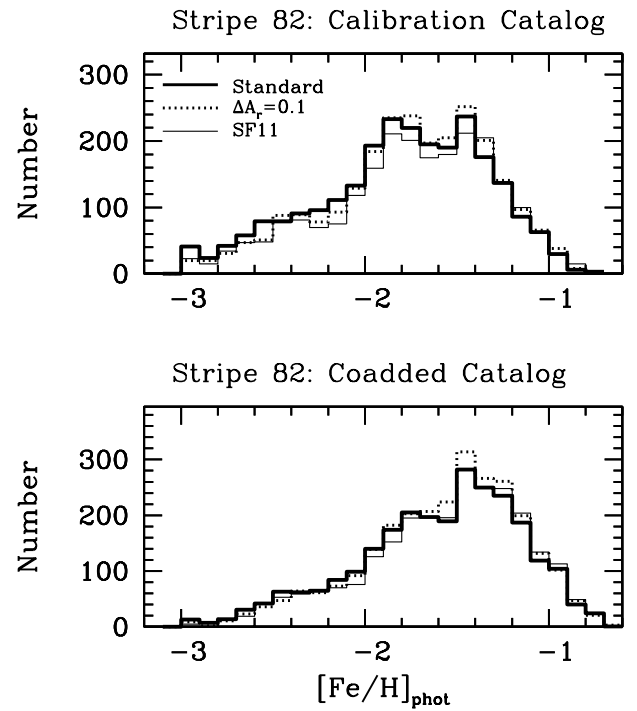}
\caption{Photometric MDFs of the halo. The black solid histogram shows a MDF
using our adopted extinctions and extinction coefficients. The dotted histogram is a MDF
assuming $10\%$ smaller extinction values; the thin solid histogram represents
a case assuming extinction coefficients from \citet{schlafly:11}. The top panel shows the MDF
from the calibration catalog and the bottom panel shows the MDF from the coadded catalog.
The photometric MDFs are complete for [Fe/H]$_{\rm true}\leq-1.2$.
\label{fig:mdf}}
\end{figure}

Figure~\ref{fig:mdf} shows our {\it in situ} observed MDFs from the calibration
(top panel) and the coadded catalogs (bottom panel), respectively, including
only stars satisfying the selection criteria described above.
\citet{juric:08} found that both thick-disk and halo stars have approximately
the same number density at $3$~kpc above the Galactic plane. In order to minimize the
contribution from thick-disk interlopers in our halo sample, we imposed a
heliocentric distance limit of greater than $5$~kpc and Galactic latitudes $|b|
> 35\arcdeg$, which correspond to a minimum vertical distance $|Z| = 2.9$~kpc
from the Galactic plane at the low-latitude limit.
As noted above, we imposed a sample cut at ${\rm [Fe/H]_{\rm true}} = -1.2$,
where ${\rm [Fe/H]_{\rm true}}$ indicates a true metallicity value.
The great majority of thick-disk stars having metallicities above this limit
would have been excluded from our photometric sample selection.
Nevertheless, it remains
possible that some metal-weak thick-disk stars \citep[][and references
therein]{chiba:00,carollo:10} have entered our sample, at least near the
low-latitude limit. Note that some stars near ${\rm [Fe/H]_{\rm true}} = -1.2$
could have photometric metallicities greater than ${\rm [Fe/H]_{\rm phot}} =
-1.2$, because of non-zero photometric errors.

The thick solid histograms in Figure~\ref{fig:mdf} show our derived photometric
MDFs, constructed using the YREC-based extinction coefficients in \S~\ref{sec:param};
the dotted lines represent the case with $10\%$ smaller extinction values than in
\citet{schlegel:98}. The thin solid histograms are those based on extinction
coefficients from \citet{schlafly:11}. Although their extinction
coefficients are $\sim20\%$ smaller than our default YREC-based values, the
difference has only a moderate effect on the resulting MDFs; the overall shape of
the MDFs does not change significantly, because of the small level of foreground
dust toward these high Galactic latitude stars. 

The photometric MDFs are insensitive to our adopted [$\alpha$/Fe] and age scheme
(\S~\ref{sec:isochrone}). Even if we had taken a different [$\alpha$/Fe] or age
relation as a function of [Fe/H], model colors would still be forced to match
the same set of cluster data from our color calibration procedure
(\S~\ref{sec:calibration}). The resulting color-[Fe/H] relation would then be
essentially unaffected by this change. Although intrinsically broad
distributions in [$\alpha$/Fe] and/or age at each [Fe/H] could result in
additional spreading of the observed MDFs, we neglect their effects in
the following discussion.

\begin{figure}
\epsscale{1.15}
\plotone{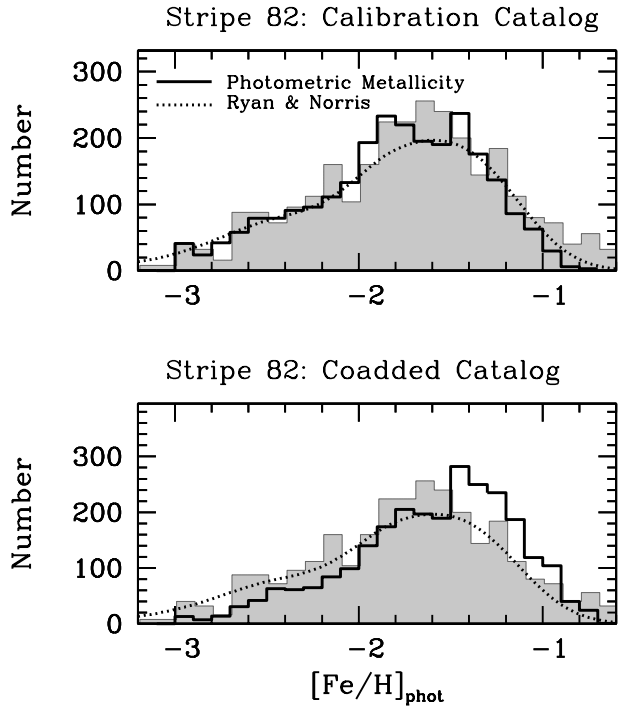}
\caption{Photometric MDFs of the halo (black solid histograms). The top panel shows the MDF
from the calibration catalog and the bottom panel shows the MDF from the coadded
catalog. Our MDFs are complete for [Fe/H]$_{\rm true}\leq-1.2$. The overlaid gray shaded
histograms are MDFs for the kinematically-selected local halo stars from
\citet{ryan:91b}, for which the sample size has been multiplied by a factor of eight, in
order to facilitate comparison with the photometric MDFs. The dotted line is the
MDF from the same set of stars \citep{ryan:91b}, but after convolving it
with the photometric convolution kernels (Figure~\ref{fig:fehprofile}), in order to simulate
the effects of photometric errors and unresolved binaries and/or blends (see text).
\label{fig:ryan}}
\end{figure}

Figure~\ref{fig:ryan} compares our photometric MDFs with the spectroscopic MDF
obtained by \citet[][gray shaded]{ryan:91b}, which is based on a
kinematically-selected sample of $372$ {\it local} halo subdwarfs ($d_{\rm helio}
\la 400$~pc). We have multiplied their sample by a factor of eight to
directly compare the shapes of MDFs. The mean [Fe/H] value in the \citet{ryan:91b}
sample is $-1.80$ over $-3 < {\rm [Fe/H]} < -1$, with a standard deviation of
$0.46$~dex. Note that, according to these authors, their sample is incomplete
above [Fe/H]$=-1$.

The abundance measurements in \citet{ryan:91b} are accurate to $\sigma\sim0.2$~dex,
while our photometric metallicity estimates have larger dispersions and asymmetric
error distributions. In order
to perform an ``apples-to-apples'' comparison with our photometric MDFs,
we convolved the MDF of \citet{ryan:91b} with the convolution kernels in
Figure~\ref{fig:fehprofile}. We binned their data at [Fe/H]$<-1$ with intervals
of $\Delta {\rm [Fe/H]} = 0.1$ dex, and applied the convolution kernels to each bin,
ignoring spectroscopic [Fe/H] errors in \citeauthor{ryan:91b}. The resulting MDF is
shown (with an arbitrary normalization) as a dotted line in Figure~\ref{fig:ryan},
which includes the effects of both photometric errors and unresolved binaries and/or
blending, as in our photometric MDFs. The \citet{ryan:91b} MDFs are similar in
their overall shape, both before and after the convolution is applied, and both
match our {\it in situ} MDFs rather well.

As shown in Figure~\ref{fig:fehprofile}, there is a systematic shift in
the metallicity scale in the coadded catalog. A Kolmogorov-Smirnov (K-S) test
rejects the null hypothesis that the MDF of \citet{ryan:91b} and that of
the coadded catalog ({\it bottom} panel) are drawn from the same parent
population at significant levels ($p < 0.001$). On the other hand, the null
hypothesis that the \citeauthor{ryan:91b} MDF and that of the calibration
catalog ({\it top} panel) are drawn from the same parent population cannot be
rejected ($p = 0.364$); in other words, their MDFs are statistically similar
with each other. Photometric zero-point shifts may be responsible for the
systematically higher photometric [Fe/H] estimates in the coadded catalog
(\S~\ref{sec:bias} and see below in \S~\ref{sec:twocomp}).

Perhaps the overall good agreement between our photometric MDFs and the
spectroscopic \citet{ryan:91b} MDFs should not be so surprising, given the
similar range of Galactocentric distances of these two samples. The
\citet{ryan:91b} stars occupy a local volume (although their orbits extend to
much larger distances), at roughly $8$~kpc away from the Galactic center. Our
sample stars in Stripe~82 are found at present distances within $7$--$14$~kpc
from the Galactic center, as shown in Figure~\ref{fig:gcdist}, although they are
much farther away from the Galactic disk than the \citet{ryan:91b} sample stars.
It is worth keeping in mind that, although we do not presently have full space
motions for the stars in our photometric MDFs, they surely contain many stars
with orbits that take them far outside the local volume. Both samples of stars
can be thought of as local ``snapshots'' of a volume in the relatively nearby
halo.

\subsection{Deconvolution of the Halo Metallicity Distribution}\label{sec:deconvolution}

We now describe our attempts to recover the ``true'' halo MDF, using experiments
that consider different ways for understanding the nature of the underlying
[Fe/H] distributions. This requires confronting the possibility that more than
one parent stellar population may be required, as discussed below.

We began by deconvolving the observed MDFs of the halo using a single
[Fe/H]$_{\rm phot}$ profile (one of the kernels in Figure~\ref{fig:fehprofile}),
where a single [Fe/H]$_{\rm true}$ is assumed, without an intrinsic
dispersion. We found that none of these profiles are able to match the
overall shape of the MDFs, immediately suggesting that either the hypothesis of
a small dispersion in [Fe/H]$_{\rm true}$ should be abandoned, and/or
at least two sub-components are required to properly account for the overall
shape of the MDF. We consider each of these two cases in turn below.

\subsubsection{A One-Component Model}

\begin{figure}
\epsscale{1.15}
\plotone{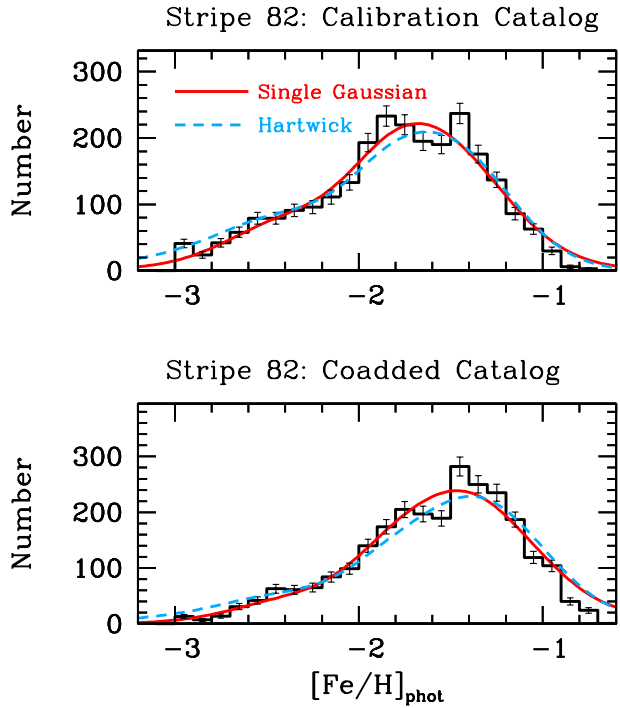}
\caption{Deconvolution of photometric MDFs of the halo using a single Gaussian
[Fe/H] distribution. Histograms are observed MDFs from the calibration
({\it top}) and the coadded catalog ({\it bottom}), respectively.
The error bars represent $\pm1\sigma$ Poisson errors. The solid red line shows
an error-convolved [Fe/H]$_{\rm phot}$ distribution from the deconvolution kernels
in Figure~\ref{fig:fehprofile}, with a peak of the underlying [Fe/H] distribution at
[Fe/H]$=-1.80$ for the calibration catalog ({\it top}), and a peak at [Fe/H]$=-1.55$
for the coadded catalog ({\it bottom}), respectively, both with dispersions of $0.4$~dex.
The dashed blue line is the best fitting simple chemical evolution model from
\citet{hartwick:76}, after applying the deconvolution kernels.
\label{fig:deconv.single}}
\end{figure}

Figure~\ref{fig:deconv.single} shows the fitting results for the observed
photometric MDFs, assuming that the metallicity of the underlying halo
population can be well-described by a single Gaussian distribution with a
dispersion in metallicity. The solid red line in Figure~\ref{fig:deconv.single}
shows the resulting fit, obtained using the non-linear least squares fitting
routine MPFIT \citep{markwardt:09} over $-2.8 < {\rm [Fe/H]_{\rm phot}} < -1.0$,
after application of the convolution kernels (Figure~\ref{fig:fehprofile}). As
described earlier, each of these simulated profiles includes the effects of
photometric errors ($\sigma_{g, r,i} = 0.02$ mag, $\sigma_{u,z} = 0.03$ mag) and
a $50\%$ unresolved binary fraction and/or blends with the M35 mass function for
secondaries. Note that these kernels are not Gaussian functions, nor are they
the products of a chemical evolution model.

For the calibration catalog (top panel), we found that a best-fitting Gaussian
has a peak at [Fe/H]$_{\rm true} = -1.80\pm0.02$ with $\sigma_{\rm [Fe/H]}=0.41\pm0.03$~dex.
The reduced $\chi^2$ value ($\chi^2_\nu$) of the fit is $2.0$ for $15$ degrees
of freedom, assuming Poisson errors. The errors in the above parameters
were scaled based on the $\chi^2_\nu$ of the fit.
The estimated peak [Fe/H] and the dispersion are similar to those obtained for the
\citet{ryan:91b} sample, as expected from the similar MDF shape from these two
studies. For the coadded catalog (bottom panel), we found a best-fitting
Gaussian with a peak at [Fe/H]$_{\rm true}=-1.55\pm0.03$ and
$\sigma_{\rm [Fe/H]}=0.43\pm0.04$~dex, with $\chi^2_\nu=2.7$.

For an additional test, we assumed that the photometric MDF is shaped primarily
by large photometric errors, even though the underlying [Fe/H] distribution is
single-peaked at around [Fe/H]$=-1.6$. We found marginal agreement with the
observed MDFs only if the size of photometric errors were underestimated by a
factor of two (i.e., $\sigma_{g,r,i} \approx 0.04$ mag, $\sigma_{u,z} \approx
0.06$ mag) for both of the Stripe~82 catalogs. However, we consider it unlikely
that the errors have been underestimated by this much. This hypothesis is also
inconsistent with the intrinsically wide range of spectroscopic [Fe/H]
determinations from \citet{ryan:91b}. 

The dashed blue line in Figure~\ref{fig:deconv.single} shows the best-fitting
simple chemical evolution model \citep{hartwick:76}. We used the mass-loss
modified version as in \citet{ryan:91b}, who found an excellent match of
this model to their MDF. The model MDF is characterized by a single parameter,
the effective yield ($y_{\rm eff}$), which relates to the mass of ejected metals
relative to the mass locked in stars. The simple mass-loss modified model
adopts instantaneous recycling and mixing of metal products in a leaky
box, and further assumes a zero initial metallicity, constant initial mass
function, and a fixed effective yield. We utilized [Fe/H] in this model as a surrogate
for the metallicity. After convolving the model MDF with the deconvolution kernels
(Figure~\ref{fig:fehprofile}), we found $\log_{10} y_{\rm eff}=-1.65\pm0.02$
($\chi^2_\nu=2.6$ for 15 degrees of freedom) for the calibration catalog, and
$\log_{10} y_{\rm eff}=-1.37\pm0.04$ ($\chi^2_\nu=5.0$) for the coadded catalog,
which simply correspond to the peak of the MDF. The goodness of the fit to the
calibration catalog data is comparable to that using the single Gaussian fit,
and the resulting effective yield is close to what \citet{ryan:91b} obtained
from their MDF ($\log_{10} y_{\rm eff}=-1.6$).

\subsubsection{A Two-Component Model}\label{sec:twocomp}

Although a single-peak Gaussian [Fe/H] distribution describes the observed
photometric MDF rather well, it is not a unique solution. Here we consider a
two-peak Gaussian [Fe/H] distribution fit to the Stripe~82 MDFs.

\begin{figure}
\epsscale{1.15}
\plotone{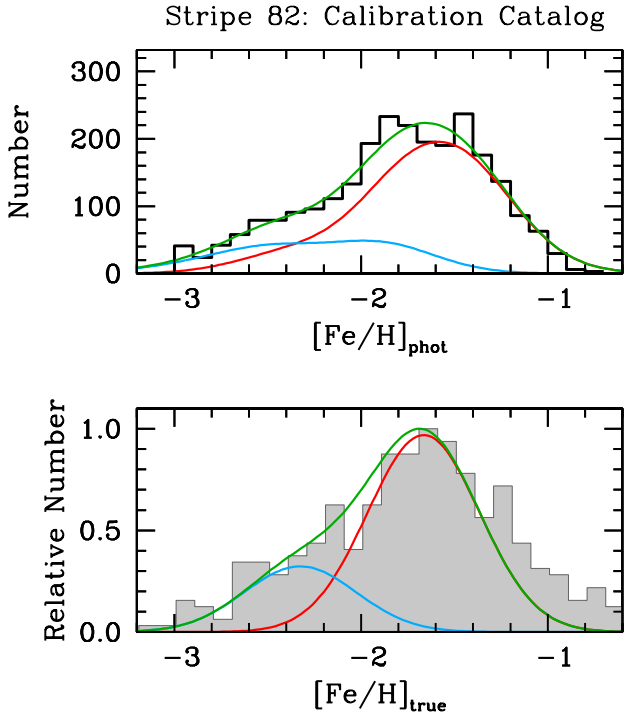}
\caption{{\it Top:} Deconvolution of the photometric MDF of the halo from
the calibration catalog, using a double Gaussian [Fe/H] distribution.
Convolution kernels (Figure~\ref{fig:fehprofile}) are applied to the
underlying [Fe/H] distribution to simulate effects of photometric errors and
unresolved binaries and/or blends in the sample. The red and blue curves show
a convolved [Fe/H]$_{\rm phot}$ distribution from each component of the
double Gaussian distribution. The green curve is the sum of these contributions.
The standard deviation of the Gaussians are set to $\sigma_{\rm [Fe/H]}=0.30$~dex.
{\it Bottom:} The true underlying [Fe/H] distribution, as derived from the
[Fe/H]$_{\rm phot}$ deconvolution shown in the top panel. The red and blue curves are
the underlying [Fe/H]$_{\rm true}$ distributions, with peaks at [Fe/H]$=-1.67$ and
$-2.33$, and the green curve is the sum of these two. The gray histogram shows
the observed [Fe/H] distribution from \citet{ryan:91b}, normalized to the peak
of the distribution.
\label{fig:deconv.calib}}
\end{figure}

The red and blue curves in the top panel of Figure~\ref{fig:deconv.calib} show
the best matching pair of simulated profiles to the calibration catalog MDF,
searched using the MPFIT routine over $-2.8 < {\rm [Fe/H]_{\rm phot}} < -1.0$.
In this fitting exercise, we fixed the dispersion of the underlying
[Fe/H] distributions to $\sigma_{\rm [Fe/H]}=0.30$~dex for both Gaussians.
The green curve is the sum of these individual components, which exhibits
an excellent fit to the observed profile ($\chi^2_\nu=1.9$ for $14$ degrees of
freedom). The underlying true [Fe/H] distribution for each of these curves is
shown in the bottom panel, which exhibits peaks at [Fe/H]$_{\rm true}=-1.67\pm0.08$
and $-2.33\pm0.30$, respectively.

The bottom panel in Figure~\ref{fig:deconv.calib} also shows the spectroscopic
MDF from \citet[][gray histogram]{ryan:91b}. Our deconvolved [Fe/H] distribution
matches their observed MDF well on the metal-poor side. Their sample is
contaminated by disk stars above [Fe/H]$\approx-1$, while our photometry sample
selection excludes metal-rich stars with [Fe/H]$_{\rm true}>-1.2$
(\S~\ref{sec:bias}).

The fractional contribution of the low-metallicity component with a peak at
[Fe/H]$_{\rm true}=-2.33$ (the area under the blue curve in the top panel of
Figure~\ref{fig:deconv.calib}) to the entire halo sample (the area under the
green curve in the top panel) is $24\%$, with a clearly strong dependence on
metallicity. Below [Fe/H]$=-2.0$, the contribution from the low-metallicity
component is $52\%$ of the total numbers of halo stars in this metallicity
regime. Above [Fe/H]$=-1.5$, the high-metallicity component (the red curve
with a peak at [Fe/H]$_{\rm true}=-1.67$) contributes $96\%$ of the total numbers
of halo stars.

The relative fraction of the low-metallicity component depends on our adopted
value for the dispersion of the underlying [Fe/H] distribution. In the above
exercise, we assumed $\sigma_{\rm [Fe/H]}=0.30$~dex for each component. However,
the contribution from the low [Fe/H] component becomes as high as $42\%$ if
the standard deviation is set to $\sigma_{\rm [Fe/H]}=0.20$~dex. On the other
hand, fixing the value to $\sigma_{\rm [Fe/H]}=0.4$~dex leads to a similar
result as in the previous section, which requires only one component in the fit.
If $\sigma_{\rm [Fe/H]}=0.2$~dex is a reasonable lower limit to the dispersion of
each of the underlying [Fe/H] components, we can only set an upper limit on the
fractional contribution of the low-metallicity component, at $\sim40\%$--$50\%$.

To investigate this further, we have also performed two-component deconvolutions
leaving each component's dispersion as a free parameter. For this, we use
the \emph{extreme deconvolution} \citep{bovy:11} technique, which fits all parameters of
arbitrary numbers of Gaussian components to data with individual uncertainties,
and which has been extended to handle non-Gaussian uncertainty kernels, such as
those in \S~\ref{sec:artificial}. The best two-component fit (with $\Delta \chi^2 = 6.5$
compared to the fixed  $\sigma_{\mathrm{[Fe/H]}} = 0.3$ dex fit above) consists
of a $69\%$ contribution from a low-metallicity component ([Fe/H]$ = -2$) with
$\sigma_{\mathrm{[Fe/H]}} = 0.35$ dex and a higher-metallicity component
([Fe/H]$ = -1.4$). However, the latter has an unrealistically small dispersion of
only  $\sigma_{\mathrm{[Fe/H]}} = 0.07$ dex. Fitting instead with a weak prior
on the dispersion (a chi-squared distribution with 2 degrees of freedom and
a mean of 0.3 dex) yields a slightly worse fit ($\Delta \chi^2 = 2$ with respect
to the best fit), with almost equal contributions from a low-metallicity (53\%,
mean [Fe/H]$ = -2.14$,  $\sigma_{\mathrm{[Fe/H]}} = 0.3$ dex) and a higher-metallicity
(mean [Fe/H]$ = -1.5$,  $\sigma_{\mathrm{[Fe/H]}} = 0.2$ dex) component. These
results further show that the parameters of a two-component fit are only poorly
constrained on the basis of the photometric-metallicity data alone.

An even stronger constraint on the fraction of the low-metallicity halo
component can be obtained if our {\it deconvolved} MDF is forced to match
the shape of the \citet{ryan:91b} MDF. As shown in the bottom panel of
Figure~\ref{fig:deconv.calib}, our assumed [Fe/H] dispersion of $0.3$~dex
results in a good agreement of our MDF with that of the Ryan \& Norris sample
(gray shaded). However, the agreement breaks down when a smaller (or larger) value
is assumed for a dispersion of the individual underlying [Fe/H] distributions; a
smaller dispersion leads to an enhanced contribution from the metal-poor
component, resulting in an overestimated number of metal-poor stars with respect
to the Ryan \& Norris MDF. From the above considerations, we find that
individual components of the underlying MDF with $\sim0.25$--$0.32$~dex
dispersion in [Fe/H] provide a reasonably good fit to the \citet{ryan:91b} MDF;
in other words, the \citeauthor{ryan:91b} MDF is also well fit by double Gaussians
with these dispersion values. The fractions of the low-metallicity component of
the halo corresponding to these values are $\sim20\%$--$35\%$ of the entire halo
population.

\begin{figure}
\epsscale{1.15}
\plotone{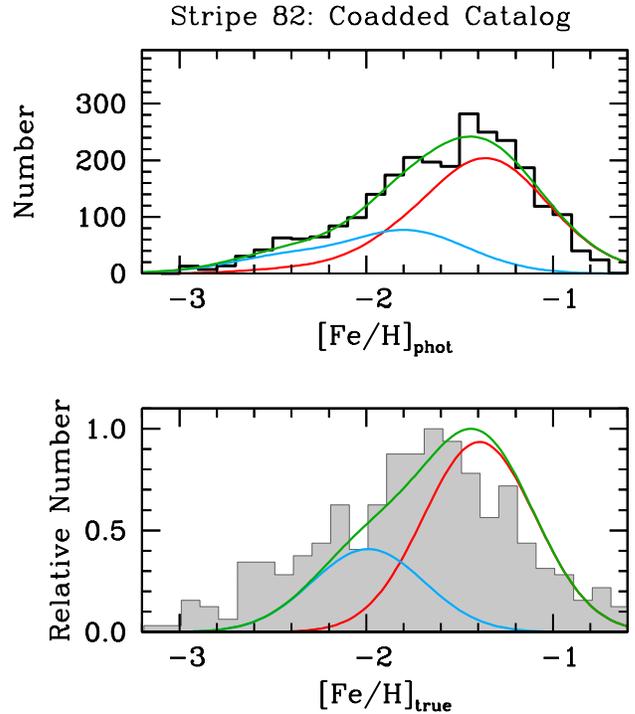}
\caption{Same as in Figure~\ref{fig:deconv.calib}, but for the coadded catalog.
The red and blue curves in the bottom panel show the underlying [Fe/H]$_{\rm true}$
distributions with peaks at [Fe/H]$=-1.39$ and $-1.99$, respectively.
\label{fig:deconv.coadd}}
\end{figure}

In the case of the coadded catalog (Figure~\ref{fig:deconv.coadd}), the overall
metallicities are shifted toward higher [Fe/H] values; peak [Fe/H]$_{\rm true}$
values of the two components are ${\rm [Fe/H]}=-1.39\pm0.12$ and $-1.99\pm0.23$,
respectively, with a total $\chi^2_\nu$ value of $2.6$.
The shift in [Fe/H] is expected from the photometric zero-point difference
between the two Stripe~82 catalogs (\S~\ref{sec:bias}), where the larger
(fainter) $u$-band magnitudes in the coadded catalog produce higher photometric
metallicities than in the calibration catalog. The contribution of the
metal-poor component is comparable ($30\%$) to that derived from the
calibration catalog ($24\%$).

A strong disagreement with the \citet{ryan:91b} MDF is evident in the lower panel,
where the deconvolved MDF shows systematically higher metallicities. If we expect
both MDFs to be the same at similar Galactocentric distances, this would indicate
that the MDF in Figure~\ref{fig:deconv.coadd} has been shifted toward higher
[Fe/H] due to photometric zero-point errors in the coadded catalog (\S~\ref{sec:bias}).
Some of the stars in Stripe~82 have spectroscopic estimates from the SSPP,
and were included in the comparison of the spectroscopic metallicities with the
photometric determinations (\S~\ref{sec:sspp}). However, there were not a sufficient
number of stars satisfying both the spectroscopic (\S~\ref{sec:sspp}) and photometric
(\S~\ref{sec:bias}) selection criteria for a meaningful comparison to verify which
photometric catalog provides a more consistent result with the SSPP estimates.

Our results illustrate a simple limiting case, where one could imagine that the
stellar halo formed out of individual subhalos which, in the mean, were
pre-enriched in metals that peak roughly at the observed locations.
Nevertheless, the fact that the observed photometric [Fe/H] distribution is well
described by primarily two metallicity peaks at [Fe/H]$\sim-1.7$ and $-2.3$
is consistent with the argument by \citet{carollo:07,carollo:10} that our Milky Way
stellar halo is a superposition of at least two components, the inner and outer
halos, that are distinct in metallicity, kinematics, and spatial distributions.
The apparent duality of the Galactic halo system is further supported by a
limited kinematic analysis of stars in our photometric sample, as described
below (\S~\ref{sec:kinematics}). It should be kept in mind that our description
of these two populations, following Carollo et al., assumes that they are
``smooth'' distributions, even though it is recognized that there exist
numerous, possibly even locally dominant, examples of substructure (such as
stellar streams and overdensities) present as well.

From the above tests, we conclude that we do not have sufficiently strong
constraints, from the observed photometric MDFs alone, in order to discriminate
between the hypothesis of
the duality of the halo (Figures~\ref{fig:deconv.calib}--\ref{fig:deconv.coadd})
and the single stellar population (Figure~\ref{fig:deconv.single}) model. We
also cannot preclude the possibility that the halo comprises more than two
sub-components, because of the limited number of constraints presented by the
observed MDFs to perform multi-component fitting. However, we show below that
the degeneracy in this solution can be at least partially removed by combining
independent information from the available kinematics of our sample stars, as
well as for those in the
\citet{ryan:91b} sample.

\subsection{Metallicity Correlations with Kinematics}\label{sec:kinematics}

Most of the stars in our Stripe~82 photometric sample do not have full velocity
information available. However, at high Galactic latitudes, proper-motion
measurements alone can provide useful constraints on the $U$
(towards the Galactic center) and $V$ (in the direction of Galactic disk
rotation) velocity components, which are sufficient to constrain the motions of
stars parallel to the Galactic plane.

We searched for proper-motion measurements in SDSS \citep{munn:04}, with the
proper-motion flags set to the recommended values: ${\tt match} = 1$,
${\tt nFit} = 6$ and ${\tt dist22} > 7$, ${\tt sigRa} < 525$, and ${\tt sigDec}
< 525$ (taking into account the correction described by Munn et al. 2008). We
then combined proper-motion measurements with our distance estimates to compute
{\it approximate} rotation velocities\footnote{Below we use $V_\phi$ to represent
a rotational velocity computed using the full kinematic information, to distinguish
it from $v_\phi$.} ($v_\phi$) in the Galactocentric cylindrical system \citep[see][]{bond:10}.

\begin{figure}
\epsscale{1.15}
\plotone{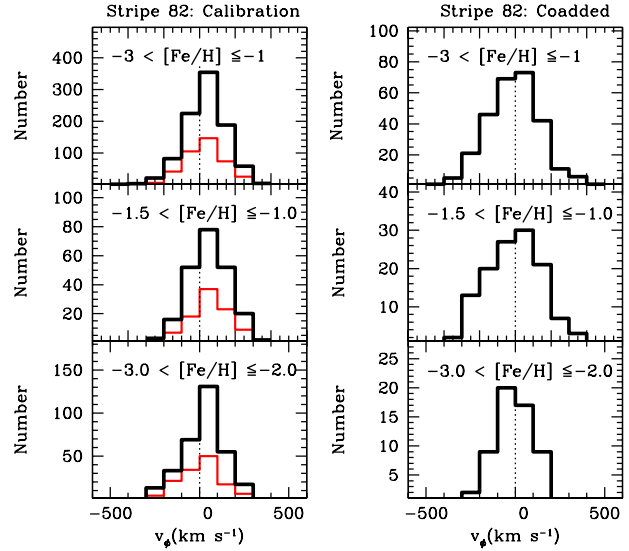}
\caption{Distribution of approximate rotation velocity in the Galactocentric cylindrical
system ($v_\phi$) for the calibration ({\it left panels}) and the coadded ({\it right panels}) catalogs.
The thick black and thin red histograms are those based on stars at $|b| > 45\arcdeg$ and
$|b| > 60\arcdeg$, respectively. The top to bottom panels show the sample in
different metallicity bins.  A vertical dotted line is shown at zero rotation.
\label{fig:ppmhist}}
\end{figure}

Figure~\ref{fig:ppmhist} shows the approximate velocity distribution in $v_\phi$ for the
calibration and coadded catalogs, respectively. The top panels show the
full velocity distribution of all stars in each catalog, while the middle and bottom
panels show subsets of these in the metal-rich ($-1.5 < {\rm [Fe/H]} \leq -1.0$;
middle panel) and in the metal-poor ($-3.0 < {\rm [Fe/H]} \leq -2.0$; bottom
panel) regimes, respectively. Inspection of the distributions in these two
metallicity ranges indicates that they are not identical, as might be expected
for a single halo population. The thick black histograms show the $v_\phi$ distributions
of stars at $|b| > 45\arcdeg$; red histograms show the more restricted case with
$|b| > 60\arcdeg$ (the restriction to higher latitude provides a better
approximation for the rotational velocity). Comparisons between the black
and red histograms show that the
latitudinal restrictions only mildly change the overall distributions in all
metallicity bins. For the calibration catalog, a K-S test
rejects the null hypothesis that the samples in the two different metallicity
ranges are drawn from the same parent population at significant levels ($p = 0.019$ for
$|b| > 45\arcdeg$; $p = 0.003$ for $|b| > 60\arcdeg$).  For the coadded catalog,
the K-S test is unable to reject this hypothesis ($p = 0.167$ for $|b| >
45\arcdeg$; $p = 0.297$ for $|b| > 60\arcdeg$). Although not a strong effect,
there exist greater fractions of more metal-poor halo stars in retrograde, rather
than prograde, rotation for both catalogs.

\begin{figure}
\epsscale{1.15}
\plotone{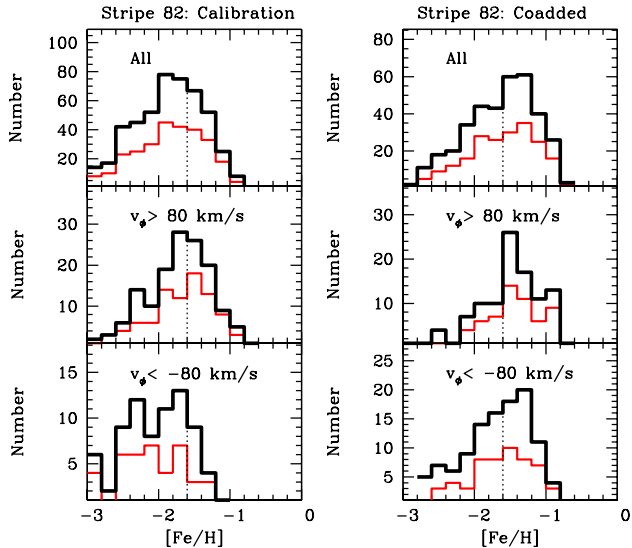}
\caption{MDFs in different bins of approximate rotational velocity in the
Galactocentric cylindrical system ($v_\phi$) for the calibration ({\it left panels}) and the
coadded ({\it right panels}) catalogs. The thick black and thin red histograms are those based on
stars at $|b| > 45\arcdeg$ and $|b| > 60\arcdeg$, respectively.  A vertical
dotted line at [Fe/H] $= -1.6$ is shown for reference.
\label{fig:ksplit}}
\end{figure}

A more striking difference is obtained when the MDFs are considered for 
different $v_\phi$ ranges, as shown in Figure~\ref{fig:ksplit}. As before, the black and
red histograms indicate the two samples at different Galactic latitude ranges.
The middle panels show the MDFs for stars with high prograde
rotation ($v_\phi > 80$~km/s); bottom panels show those of stars with extreme
retrograde rotation ($v_\phi < -80$~km/s).  Note that the $\pm80$~km/s limit
was set in order to maximize the numbers of stars in each split, so as to enable
statistically meaningful interpretations. 

Inspection of Figure~\ref{fig:ksplit} indicates that the MDFs for stars in
retrograde motion are shifted toward lower metallicities than the MDFs for those
in prograde rotation. For the calibration catalog, whether the samples are
restricted to $|b| > 45\arcdeg$ or $|b| > 60\arcdeg$, a K-S test rejects the
null hypothesis that the two samples for different ranges of $v_\phi$ are drawn
from the same parent population at highly significant levels ($p = 0.0001$ for
$|b| > 45\arcdeg$; $p = 0.0002$ for $|b| > 60\arcdeg$). For the coadded catalog,
the lower-latitude cut results in a significant rejection ($p = 0.026$ for $|b| >
45\arcdeg$); no rejection can be made for $|b| > 60\arcdeg$ ($p = 0.257$). We
conclude that greater fractions of low-metallicity halo stars exist in
retrograde, rather than prograde, rotation. 

\begin{figure}
\epsscale{1.15}
\plotone{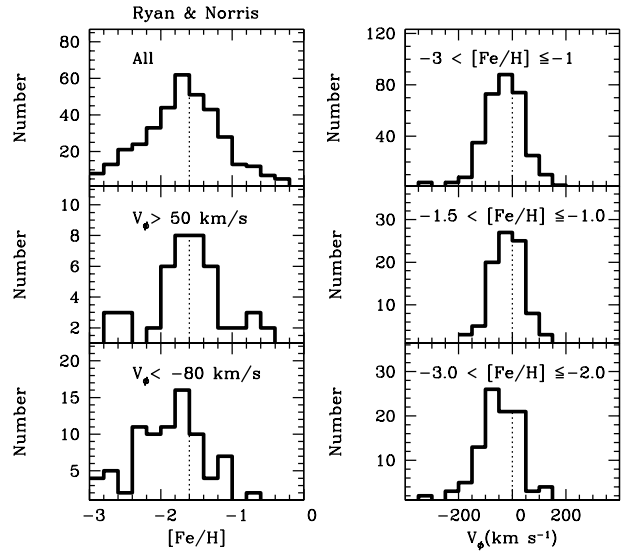}
\caption{Same as in Figures~\ref{fig:ppmhist}-\ref{fig:ksplit}, but for the local
kinematically-selected sample of \citet{ryan:91b}. Note that we distinguish $V_\phi$
from $v_\phi$ to indicate the fact that the rotation component for the
\citeauthor{ryan:91b} sample is obtained from full space motions. The vertical dotted line is
shown in the left panels for [Fe/H] $= -1.6$, and at zero rotation in the right
panels, for reference.
\label{fig:ksplit2}}
\end{figure}

Inspired by these results, we decided to check how the MDFs of the spectroscopic
\citet{ryan:91b} sample vary for prograde vs.\ retrograde rotation, and how the
$V_\phi$ distributions vary for higher- and lower-metallicity divisions. Note
that we distinguish $V_\phi$ from $v_\phi$ to indicate the fact that the
rotation component for the Ryan \& Norris sample is obtained from full space
motions \citep{ryan:91a}. Figure~\ref{fig:ksplit2} shows the [Fe/H] distributions 
(left panels) and $V_\phi$ distributions (right panels) of their local
kinematically-selected sample, with the same limits as on $v_\phi$ and [Fe/H] in
Figures~\ref{fig:ppmhist}-\ref{fig:ksplit}, except in the middle left panel,
where we set $V_\phi > 50$~km/s to include a sufficient number of stars. There
exist clear differences in these distributions. A K-S test indicates that the
null hypothesis that the samples are drawn from the same parent population is
rejected at significant levels: $p = 0.019$ for the different ranges in $V_\phi$
and $p= 0.021$ for the different ranges in [Fe/H]. Both divisions indicate that
greater fractions of metal-poor halo stars are found in retrograde, rather than
prograde, rotation.

\input{tab3.tex}

Table~\ref{tab:tab3} summarizes the average [Fe/H] values (and errors in the
mean) for the kinematically-divided samples considered above. For the
calibration and coadded catalogs, the mean [Fe/H] is between $-1.7$ and $-1.5$
for stars in highly prograde motion ($v_\phi\ > 80$~km/s). For the
calibration catalog, the mean [Fe/H] for the highly retrograde components
($v_\phi < -80$~km/s) is $\sim 0.3-0.4$~dex lower than for the metal-rich
counterparts; the difference is $\sim0.2$~dex for the coadded catalog. The
difference in the mean [Fe/H] between the highly retrograde component and the
highly prograde component in the \citet{ryan:91b} sample is 0.3 dex. Taken as a
whole, these simple kinematic divisions indicate that a single stellar
population is inadequate to describe the behavior of the higher and
lower-metallicity halo stars for all three samples.

We caution that the observed kinematic-metallicity correlation can also be
produced by (sub)giant contamination in our sample \citep[e.g.,][]{ryan:89}.
Unrecognized giants and subgiants are brighter than dwarfs, so their distances
and velocities will be underestimated, while their photometric metallicities
are higher than those for dwarfs (Figure~\ref{fig:ugr}). We used the photometry
of M13 \citep{an:08} to perform the same suite of simple Galactic halo models
as in \S~\ref{sec:giants}, and found that (sub)giants have systematically higher
photometric metallicities, and higher $v_\phi$, than dwarfs. The overall
contamination rate of (sub)giants in the sample is estimated around $15\%$,
but the fraction increases to $\sim40\%$ when high-metallicity stars
($-1.5 < {\rm [Fe/H]} < -1.0$) are considered. However, it is less likely that
the observed kinematic-metallicity correlation is caused entirely by the
(sub)giant contamination, because the \citeauthor{ryan:91b} sample,
which is independent from our photometric sample and has different selection
criteria, also exhibits the same behavior (Figure~\ref{fig:ksplit2}). Further
investigation is needed on the effect of (sub)giant contamination.

\section{Summary and Discussion}\label{sec:discussion}

We have used a new, empirically calibrated set of stellar isochrones to derive
distances, temperatures, and metallicities for individual stars in SDSS
Stripe~82 using $ugriz$ photometry. These estimates have been validated by
comparisons with the IRFM temperature scale and SDSS/SEGUE medium-resolution
spectroscopic values. Based on unbiased photometric samples of main-sequence
Stripe~82 stars in a relatively narrow mass range, we have constructed an
{\it in situ} MDF of the Galactic halo in the distance range $5-8$ kpc from
the Sun, which is similar in shape to that of the local kinematically-selected
subdwarfs in \citet{ryan:91b}.
This MDF can be adequately fit using a single Gaussian with peak at [Fe/H]
$=-1.5$ and a broad (0.4 dex) dispersion of metallicities. However, an equally
valid fit can be obtained from the use of two components with peaks at [Fe/H]
$\approx -1.7$ and $-2.3$, with smaller dispersions ($0.25$--$0.32$~dex). For an
adopted dispersion of 0.3~dex, on the order of 25\% of the stars in our local
halo sample can be associated with the low-metallicity component. For stars with
[Fe/H] $< -2.0$, this fraction increases to on the order of 50\%. For stars with
[Fe/H] $> -1.5$, essentially all of the stars are associated with the
high-metallicity component. If a smaller dispersion is adopted (0.2~dex), $\sim
40\%-50$\% of the total halo sample can be associated with the low-metallicity
component, but would make the deconvolved photometric MDF different in shape
from that of the local \citet{ryan:91b} kinematically-selected sample. Our analysis cannot
preclude the possibility that more than two sub-components could exist in the
halo.

A limited kinematic analysis of the stars in our photometric sample, based on
approximation of the rotational motions of stars about the Galactic center
(derived from available proper motions at high Galactic latitude), indicates
that the metal-poor and metal-rich subsamples exhibit different behaviors,
suggesting that the dual-component model is likely to be superior to the
single-component model. Greater fractions of metal-poor stars ([Fe/H] $< -2.0$)
in the halo possess retrograde orbits, as compared to more metal-rich stars
($-1.5 <$ [Fe/H] $< -1.0$), consistent with the claim by \citet{carollo:07,
carollo:10} that the halo comprises two overlapping systems, an inner-halo
population and an outer-halo population, with similar metallicity-kinematic
correlations as we find from our present analysis.
The observed behaviors of these halo stars are also consistent with
theoretical expectations from Milky Way-like galaxies, where inner-spheroid
stars primarily formed in the main progenitor(s) of the galaxy and exhibit net
prograde rotation, while the accreted outer-halo component exhibits a mild
degree of rotation, or in a minority of cases, a retrograde rotation
\citep{mccarthy:12}.

We have also considered an independent sample of halo stars, one that is still
in common use for describing the MDF of the Milky Way's halo. \citet{ryan:91b}
found that their derived halo MDF (based on a local kinematically-selected
sample) can be well-described by a simple chemical evolution model
\citep{hartwick:76}. Our local {\it in situ} halo MDF is
also well-fit by this model. However, there are a number of unphysical
assumptions that are made by such a model, including instantaneous recycling and
mixing of metals throughout the entire halo, as well as the use of a so-called
effective yield (by which the metals produced by individual supernovae are
reduced by an arbitrary amount, one that is not predicted by present
understanding, but acts only as a fitting parameter of the model). Such
limitations immediately call into question the applicability of these classes of
models in the context of a presumed hierarchical assembly of the Galactic halo
from individual lower-mass subhaloes. Such a model (at least when applied to the
halo as a whole) clearly presumes that the stars belong to a single population.
As was the case for our analysis of the unbiased photometric stellar halo
sample, our investigation of the rotational behavior of the Ryan \& Norris data
also does not support a single-population model.

Our current understanding on the nature of the halo provides caveats for the use
and interpretation of the observed MDFs. For example, \citet{schoerck:09} and
\citet{li:10} consider halo MDFs based on spectroscopic follow-up of candidate
metal-poor stars from the Hamburg/ESO survey. Below [Fe/H]$ = -2.5$, the MDFs of
both samples (the first comprising mainly giants, the second comprising mostly
dwarfs) appear in excellent agreement with one another, and led to the conclusion
that there existed a sharp low-metallicity cutoff (at [Fe/H]$ = -3.6$), below which
fewer than expected numbers of the lowest metallicity stars were found. For stars with
higher metallicity, these authors argue that the observed MDFs of the two samples
differ from one another due to the very different selection biases on metallicity
involved for the cooler (giants) and warmer (dwarfs) samples. However, if the halo comprises
two overlapping stellar components \citep{carollo:07,carollo:10,beers:12},
the different behaviors of the giant and dwarf samples are likely to be strongly
influenced by the degree to which the inner- and outer-halo populations contribute
to them, convolved with the clearly present selection biases, rather than being
due to the selection bias alone. The implication of the dual nature of the halo is
that the subject under discussion is {\it not} the tail of the MDF of ``the halo,''
but rather, the tail of the MDF of the outer-halo population, with clear implications
for the origin of this component. Further investigations of these connections
would be of great interest.

\acknowledgements

D.A.\ thanks Gerry Gilmore for helpful comments. We also thank Brian Yanny
and James Annis for useful discussions.
This work was supported by the Ewha Womans University Research Grant of 2010,
as well as from support provided by the National Research Foundation of Korea to
the Center for Galaxy Evolution Research. T.C.B.\ and Y.S.L.\ acknowledge partial
support of this work from grants PHY 02-16783 and PHY 08-22648: Physics
Frontiers Center/Joint Institute for Nuclear Astrophysics (JINA), awarded by the
U.S. National Science Foundation.
J.A.J.\ acknowledges support from NSF grant AST-0707948.
J.B.\ acknowledges support provided by NASA through Hubble Fellowship grant
HST-HF-51285.01 awarded by the Space Telescope Science Institute, which is
operated by the Association of Universities for Research in Astronomy,
Incorporated, under NASA contract NAS5-26555.
M.N.\ acknowledges support from NSF grant AST-1009670.

Funding for SDSS and SDSS-II has been provided by the Alfred P.\ Sloan
Foundation, the Participating Institutions, the National Science Foundation, the
U.S.\ Department of Energy, the National Aeronautics and Space Administration,
the Japanese Monbukagakusho, the Max Planck Society, and the Higher Education
Funding Council for England. The SDSS Web Site is http://www.sdss.org/.

The SDSS is managed by the Astrophysical Research Consortium for the
Participating Institutions. The Participating Institutions are the American
Museum of Natural History, Astrophysical Institute Potsdam, University of Basel,
University of Cambridge, Case Western Reserve University, University of Chicago,
Drexel University, Fermilab, the Institute for Advanced Study, the Japan
Participation Group, Johns Hopkins University, the Joint Institute for Nuclear
Astrophysics, the Kavli Institute for Particle Astrophysics and Cosmology, the
Korean Scientist Group, the Chinese Academy of Sciences (LAMOST), Los Alamos
National Laboratory, the Max-Planck-Institute for Astronomy (MPIA), the
Max-Planck-Institute for Astrophysics (MPA), New Mexico State University, Ohio
State University, University of Pittsburgh, University of Portsmouth, Princeton
University, the United States Naval Observatory, and the University of
Washington.

\end{document}

%% file: tab1.tex
\begin{deluxetable}{lrrrrrr}
\tablewidth{0pt}
\tablecaption{Photometric Zero-Point Differences\label{tab:tab1}}
\tablehead{
  \colhead{Cluster} &
  \colhead{SDSS} &
  \multicolumn{5}{c}{{\tt UberCal} - {\tt Photo}} \nl
  \cline{3-7}
  \colhead{Name} &
  \colhead{Runs} &
  \colhead{$\langle \Delta u \rangle$} &
  \colhead{$\langle \Delta g \rangle$} &
  \colhead{$\langle \Delta r \rangle$} &
  \colhead{$\langle \Delta i \rangle$} &
  \colhead{$\langle \Delta z \rangle$}
}
\startdata
M15      & $2566$ &  $ 0.020$ & $ 0.007$ & $-0.005$ & $-0.006$ & $-0.010$ \nl
M15      & $1739$ &  $ 0.008$ & $ 0.003$ & $-0.005$ & $ 0.003$ & $-0.009$ \nl
M92      & $4682$ &  $-0.037$ & $-0.013$ & $-0.010$ & $-0.015$ & $-0.034$ \nl
M92      & $5327$ &  $-0.001$ & $-0.004$ & $-0.031$ & $-0.035$ & $-0.026$ \nl
M13      & $3225$ &  $ 0.011$ & $-0.003$ & $-0.001$ & $-0.005$ & $ 0.006$ \nl
M13      & $3226$ &  $ 0.007$ & $-0.011$ & $-0.006$ & $-0.002$ & $ 0.007$ \nl
M3       & $4646$ &  $ 0.039$ & $ 0.014$ & $ 0.016$ & $ 0.010$ & $ 0.008$ \nl
M3       & $4649$ &  $-0.022$ & $ 0.011$ & $ 0.013$ & $ 0.004$ & $-0.015$ \nl
M5       & $1458$ &  $-0.002$ & $-0.005$ & $-0.001$ & $ 0.004$ & $-0.001$ \nl
M5       & $2327$ &  $-0.010$ & $-0.001$ & $-0.007$ & $ 0.008$ & $ 0.019$ \nl
M67      & $5935$ &  $-0.002$ & $ 0.001$ & $-0.002$ & $ 0.001$ & $-0.004$ \nl
M67      & $5972$ &  $ 0.004$ & $-0.001$ & $-0.001$ & $ 0.007$ & $-0.009$ \nl
NGC~6791 & $5416$ &  $ 0.003$ & $ 0.003$ & $ 0.004$ & $ 0.004$ & $ 0.008$ \nl
NGC~6791 & $5403$ &  $ 0.001$ & $ 0.001$ & $ 0.001$ & $ 0.000$ & $ 0.000$ \nl
\enddata
\end{deluxetable}

%% file: tab2.tex
\begin{deluxetable}{lccccc}
\tablewidth{0pt}
\tablecaption{Adopted Cluster Parameters\label{tab:tab2}}
\tablehead{
  \colhead{Cluster} &
  \colhead{$(m - M)_0$} &
  \colhead{$E(B - V)$} &
  \colhead{[Fe/H]} &
  \colhead{Age\tablenotemark{a}} &
  \colhead{} \nl
  \colhead{Name} &
  \colhead{(mag)} &
  \colhead{(mag)} &
  \colhead{(dex)} &
  \colhead{(Gyr)} &
  \colhead{References}
}
\startdata
M15     & $15.25$ & $0.10$ & $-2.42$ & $12.6$ & 1   \nl
M92     & $14.64$ & $0.02$ & $-2.38$ & $12.6$ & 1,2 \nl
M13     & $14.38$ & $0.02$ & $-1.60$ & $12.6$ & 1,2 \nl
M3      & $15.02$ & $0.01$ & $-1.50$ & $12.6$ & 1   \nl
M5      & $14.46$ & $0.03$ & $-1.26$ & $12.6$ & 1,2 \nl
M67     & $ 9.61$ & $0.04$ & $+0.00$ & $ 3.5$ & 3   \nl
NGC6791 & $13.02$ & $0.10$ & $+0.40$ & $10.0$ & 4   \nl
\enddata
\tablerefs{References for $(m - M)_0$, $E(B - V)$, and high resolution spectroscopic [Fe/H] values:
(1) \citet{kraft:03};
(2) Subdwarf-fitting distances from \citet{carretta:00};
(3) \citet{an:07b}, and references therein;
(4) An et~al. (2012, in preparation), and references therein.}
\tablenotetext{a}{Adopted ages of clusters.}
\end{deluxetable}

%% file: tab3.tex
\begin{deluxetable*}{llrrrr}
\tablewidth{0pt}
\tablecaption{Mean [Fe/H] of Kinematically Separated Samples\label{tab:tab3}}
\tablehead{
  \colhead{} &
  \colhead{Latitudinal} &
  \multicolumn{2}{c}{$v_\phi > 80$~km/s} &
  \multicolumn{2}{c}{$v_\phi < -80$~km/s} \nl
  \cline{3-4} \cline{5-6}
  \colhead{Sample} &
  \colhead{restriction} &
  \colhead{$\langle {\rm [Fe/H]} \rangle$} &
  \colhead{$N$} &
  \colhead{$\langle {\rm [Fe/H]} \rangle$} &
  \colhead{$N$}
}
\startdata
Stripe~82 Calibration & $|b| > 60\arcdeg$ & $-1.69\pm0.05$ & 89  & $-2.08\pm0.08$ & 41  \nl
Stripe~82 Calibration & $|b| > 45\arcdeg$ & $-1.72\pm0.04$ & 143 & $-2.05\pm0.05$ & 75  \nl
Stripe~82 Coadded     & $|b| > 60\arcdeg$ & $-1.47\pm0.05$ & 59  & $-1.63\pm0.06$ & 56  \nl
Stripe~82 Coadded     & $|b| > 45\arcdeg$ & $-1.51\pm0.04$ & 102 & $-1.69\pm0.05$ & 111 \nl
\citet{ryan:91b}\tablenotemark{a} & \nodata & $-1.59\pm0.09$\tablenotemark{b} & 47  & $-1.90\pm0.07$ & 88  \nl
\enddata
\tablenotetext{b}{Rotation component for the \citeauthor{ryan:91b} sample obtained
from full space motions ($V_\phi$) is used to divide the sample in the above table.}
\tablenotetext{b}{$V_\phi > 50$~km/s was used.}
\end{deluxetable*}